\documentclass{aa}
\usepackage[varg]{txfonts}
\usepackage{natbib}
\bibpunct{(}{)}{;}{a}{}{,}
\usepackage{footnote}
\usepackage{longtable}
\usepackage{lscape}
\usepackage{verbatim}
\usepackage{makecell}
\usepackage{multirow}
\usepackage{booktabs}
\usepackage{soul}
\usepackage{gensymb}
\usepackage{graphicx}
\usepackage{placeins}
\usepackage{ragged2e}
\usepackage{hyperref}
\usepackage{array}
\usepackage{textcomp}
\usepackage{float}
\usepackage{lastpage}
\usepackage{xcolor}

\newcommand\rurl[1]{%
  \href{http://#1}{\nolinkurl{#1}}%
}

\makeatletter
\renewcommand*\aa@pageof{, page \thepage{} of \pageref*{LastPage}}
\makeatother

\begin{document} 

\title{BlackGEM observations of compact pulsating stars}
\subtitle{Mode identification for the DOV PG 1159--035 using multi-colour photometry}

\author{P. Ranaivomanana 
          \inst{1,2} \and C. Johnston \inst{1,2,3}  \and M. Uzundag \inst{2} \and P.J. Groot \inst{1,4,5,6} \and T. Kupfer \inst{7,8} \and S. Bloemen \inst{1} \and P.M. Vreeswijk \inst{1} \and J.C.J. van Roestel \inst{9}  \and D.L.A. Pieterse \inst{1} \and J. Paice \inst{10} \and A. Kosakowski \inst{11,12} \and G. Ramsay \inst{13} \and C. Aerts \inst{1,2,14}
          } 
   \institute{
   Department of Astrophysics/IMAPP, Radboud University, P.O.Box 9010, 6500 GL Nijmegen, The Netherlands\\
   \email{princy.ranaivomanana@ru.nl}
   \and 
   Instituut voor Sterrenkunde, KU Leuven, Celestijnenlaan 200D, 3001 Leuven, Belgium \and
   Max-Planck-Institut für Astrophysik, Karl-Schwarzschild-Straße 1, 85741 Garching bei München,Germany \and
   Department of Astronomy, University of Cape Town, Private Bag X3, Rondebosch, 7701, South Africa \and South African Astronomical Observatory, P.O. Box 9, Observatory, 7935, South Africa \and  The Inter-University Institute for Data Intensive Astronomy, University of Cape Town, Private Bag X3, Rondebosch, 7701, South Africa \and
   Hamburger Sternwarte, University of Hamburg, Gojenbergsweg 112, 21029 Hamburg, Germany \and 
   Texas Tech University, Department of Physics \& Astronomy, Box 41051, 79409, Lubbock, TX, USA \and Anton Pannekoek Institute for Astronomy, University of Amsterdam, 1090 GE Amsterdam, The Netherlands \and Centre for Extragalactic Astronomy, Department of Physics, Durham University, South Road, Durham, DH1 3LE, UK 
   \and Department of Physics and Astronomy, Texas Tech University, 2500 Broadway, Lubbock, TX 79409, USA \and Department of Physics and Astronomy, University of North Carolina at Chapel Hill, Chapel Hill, NC 27599, USA  \and Armagh Observatory \& Planetarium, College Hill, Armagh, BT61 9DG, UK \and Max Planck Institute for Astronomy, Königstuhl 17, 69117 Heidelberg, Germany
    }
         
   \date{Received month day, year; accepted month day, year}

 
  \abstract
{While space-based telescopes offer unparalleled precision for asteroseismology, ground-based observations remain crucial for identifying compact pulsator candidates and enabling their pulsational study through multi-colour photometry. The BlackGEM telescope array, with its high-cadence multi-colour photometry survey, significantly enhances the detection and characterisation possibilities for compact pulsators that tend to be  much fainter than dwarfs or giant pulsators.}
{
Using BlackGEM multi-colour photometry of the hot pre-white dwarf PG\,1159--035, we demonstrate its capability to detect short, multi-periodic pulsations with amplitudes down to a few milli-magnitudes. The primary aim of this study is to establish the feasibility of pulsation mode identification in hot subdwarfs and white dwarfs via mode amplitude-ratio analysis derived from BlackGEM multi-colour observations.
}
{Pulsation frequencies were extracted from our target using iterative pre-whitening analysis. To validate our data-driven mode identification concept using multi-colour photometry, we used the well-studied hot pre-white dwarf PG\,1159--035, with previously identified pulsation modes, as a prototypical object that  served for  validation.}
{
The pre-whitening analysis using BlackGEM's standard $q$-, $u$-, and $i$-band light curves of PG\,1159--035 revealed pulsation frequencies of $\ell=1$ and $\ell=2$ modes, consistent with values obtained from the literature. Using the frequencies identified from the $q$ band, amplitudes in the $i$ and $u$ bands could be estimated. Subsequent amplitude ratio calculations resulted in discernible distributions for the $\ell=1$ and $\ell=2$ modes. The future assembly of more BlackGEM amplitude ratios for well-known white dwarfs with already identified modes will lead to density estimators suitable for identifying newly detected modes in known or as-yet-undiscovered pulsators.}
{
Our proof-of-concept study paves the way for large-scale asteroseismic analyses of optically faint compact pulsating stars using ground-based facilities, such as BlackGEM. As BlackGEM continues its observations, a substantial number of these objects will be observed as part of a regular survey, enabling a robust characterisation of their pulsation modes in the context of population studies.
}


\titlerunning{BlackGEM observations of compact pulsating stars.}
\authorrunning{Ranaivomanana et al.}
\maketitle
%

\section{Introduction}
Compact pulsating stars including hot subdwarfs, pre-white dwarfs, and white dwarfs are crucial laboratories for probing the late stages of stellar evolution and the internal structure of evolved low-mass stars, especially those originating from binary interactions. Despite decades of study, the evolutionary pathways connecting these objects, such as hot subdwarfs and pre-white dwarf stars, are not yet fully constrained \citep{Charpinet2000,Fontaine2008,Corsico2019,Lynas-Gray2021}. Their multiple pulsation frequencies and modes are crucial components to enable more detailed studies of their internal structures through asteroseismology \citep[see e.g. Chapter 6 in][]{Aerts2010}. 

Hot subdwarf B (sdB) stars are evolved, core-helium-burning objects located on the extreme horizontal branch \citep{2016PASP..128h2001H}. They originate from low- to intermediate-mass stars that have undergone helium ignition in their cores, although their precise formation channels are still a matter of debate. Plausible channels include binary interactions, enhanced mass loss during the red giant phase, or mergers of helium-core white dwarfs. A sub-set of these stars (sdBV stars) exhibit multi-periodic,  non-radial pulsations, which are manifested either as pressure ($p$), gravity ($g$), or both $p$ and $g$ modes. The latter class is referred to as hybrid pulsators \citep{Kilkenny1997,Green2003,Baran2005,Schuh2006}.

White dwarfs, on the other hand, represent the final evolutionary stage of low- and intermediate-mass stars, which account for over 95 per cent of all stars in the Milky Way \citep{Althaus2010}. A considerable proportion of them display photometric variability arising from pulsations, binarity, or transiting planetary debris, which provides a valuable set of diagnostics for probing their internal structures and their interactions with remnant planetary material. As the ultimate products of stellar evolution for the majority of low-to-intermediate-mass stars, white dwarfs may pass through one of three principal instability regions as they gradually cool (i.e. DOV, DBV, and DAV). The specific instability region a star enters depends on its atmospheric composition (see the reviews by \citealt{Fontaine2008,Winget2008,Corsico2019}).

At higher effective temperatures and lower surface gravities, the PG 1159 stars (hydrogen-deficient, carbon- and oxygen-rich pre-white dwarfs, and white dwarfs) show multi-periodic, non-radial $g$-mode pulsations. These are excited by the $\kappa$-mechanism operating in the partial ionisation zones of carbon and oxygen \citep{Starrfield1983,Bradley1996,Gautschy2005,Quirion2007}. Such pulsators, known as GW Vir stars, form a key link between post-asymptotic giant branch (post-AGB) stars and white dwarfs \citep{Werner2006,Werner2014,Corsico2019}. Of the 67 PG 1159 stars known \citep{2022A&A...658A..66W}, 24 have been confirmed as GW Vir pulsators \citep{Uzundag2021,2022MNRAS.513.2285U}, implying that roughly one-third of objects within the instability strip exhibit pulsations.

Asteroseismology relies critically on mode identification, which requires us to determine the spherical degree, $\ell$, and azimuthal order, $m$, of observed modes,  allowing each mode to be matched with its theoretical counterpart \citep{Aerts2010}. In slow rotators, these quantum numbers uniquely describe the pulsation geometry and determine which regions of the stellar interior are being probed. For bright, large-amplitude pulsators, $\ell$ and $m$ can often be inferred from frequency patterns in long, uninterrupted light curves \citep[see ][for reviews]{HekkerJCD2017,GarciaBallot2019,Aerts2021,Kurtz2022,AertsTkachenko2024}. However, compact pulsators are typically faint and require high-cadence, high signal-to-noise (S/N) light curves,  making them under-represented in space-based asteroseismic samples \citep[e.g.][]{Reed2016,Hermes2017,Ketzer2017,Baran2019,Reed2021,Corsico2022,Uzundag2022,Calcaferro2024}. Although many sdBVs and white dwarfs have been detected from space missions, only a modest sub-set of them possesses the continuous, high-quality photometry needed for robust mode identification. Consequently, ground-based observations remain essential for expanding the sample of compact pulsators suitable for asteroseismology and for identifying their modes through multi-colour photometry.

Traditional methods for photometric mode identification exploit amplitude ratios and phase differences measured in multiple filters \citep{Heynderickx1994,Dupret2003,Ramachandran2004}.
These techniques have been successfully applied to a variety of pulsating classes, including $\beta$ Cephei stars \citep{Heynderickx1994,Fritzewski2025}, $\delta$ Scuti stars \citep{Balona2001}, white dwarfs \citep{Tucker2018}, and hot subdwarfs \citep{Randall2005,Jeffery2008}. Notably, \object{Balloon 090100001} remains a benchmark sdBV case where multi-colour mode identification was independently confirmed by spectroscopy \citep{Baran2008}.

Through large-scale photometric and spectroscopic surveys, the sample of known pulsating hot compact stars has steadily increased to levels where carrying out individual follow-ups for each target is becoming increasingly difficult. It is therefore essential to develop methods where parameters, including mode identifications, can be obtained from the survey data directly, without any need for follow-up observations of each individual object. This requires high photometric precision on the survey products, as well as (at least) a three-filter strategy to enable the mode identification of non-radial pulsations. The same mode identification technique can be applied when at least two filters and radial velocities are utilised. Even if the instrument was not primarily designed for it, these requirements are met by the BlackGEM array \citep{Groot2024}.

The BlackGEM array comprises three robotic optical telescopes, with each telescope covering a field of view of 2.7 square degrees, located at the ESO La Silla Observatory, Chile. While its prototype, the MeerLICHT telescope, has been operational in Sutherland, South Africa, since 2019, BlackGEM became operational in 2023. BlackGEM's primary objective is to detect and characterise optical counterparts to gravitational wave events. In addition, the BlackGEM Fast Synoptic Survey \citep{Groot2024} aims to find short-period (ultra-)compact binaries, through a strategy where selected fields are observed for two 3-hour time stretches on subsequent nights, following a sequence of $q, u, q, i$-band cycles, with 60s integrations and an overhead of $\sim$15s per exposure to allow for read-out and filter changes. In addition, each field is observed in a single $q$-band exposure per night in the two weeks preceding and following the two-night high-cadence monitoring.

BlackGEM's unique capabilities, including contemporaneous observations in multiple filters, high-cadence observations, and a large field of view, make it in principle well-suited for asteroseismic survey studies of compact pulsators. In this pilot study, we demonstrate the feasibility of deriving mode identifications for compact pulsators using standard multi-colour survey products, focusing on the prototype DOV star PG\,1159--035. Demonstrating the efficiency of this mode identification technique on a selected known pulsator is essential for future large-scale asteroseismic analyses of compact star populations. The structure of this paper is as follows. Section~\ref{sec:data_and_method} describes the data and methods. Section~\ref{sec:result} presents the resulting frequency analysis and mode identifications. The discussion and conclusions are provided in Sect.~\ref{sec:conclusion_chp5}.

\section{Data and methods}\label{sec:data_and_method}

\subsection{Target}
In this pilot project, PG\,1159--035 (EPIC 201214472 or TIC\,35062562; \citealt{Winget1991,2021A&A...645A.117C,Oliveira2022}) was selected as our prime proof-of-concept target, as it has been previously studied, with numerous identified modes of low-degree $\ell$ available. Additionally, PG\,1159--035 has been found to have both $\ell=1$ and $\ell=2$ modes,  essential to testing the validity of our method. PG\,1159--035 is a hot pre-white dwarf pulsator \citep{McGraw1979}, discovered as a blue object with an apparent magnitude in the Johnson $B$-filter of 14.5 mag from the Palomar Green (PG) survey \citep{Green1977,Winget1985}. Its temperature was found to be $\rm 140\,000 \pm 5\,000 K$ \citep{Werner1991,Costa2008}, and it is considered the prototype of the spectroscopic class `PG1159' and the pulsating class DOV.

Extensive asteroseismic analyses of PG\,1159--35 have been carried out since the detection of 125 non-radial $g-$mode pulsations in its spectrum \citep{Winget1991}, using 24 hr coverage multi-site observations over 12 days conducted by the Whole Earth Telescope \citep[WET;][]{Nather1990}. Similar observations were subsequently conducted by WET spanning 15 days \citep{Bruvold1993,Costa2003} 

Mode identifications were most recently extracted based on Transiting Exoplanet Survey Satellite \citep[TESS;][]{2021A&A...645A.117C} and Kepler K2 \citep{Oliveira2022} observations. \citet{Oliveira2022} reported 32 and 27 frequencies, associated with $\ell = 1$ and $\ell = 2$ modes, respectively. Additionally, rotational splitting was identified in PG\,1159--035, suggesting a rotational period of 1.4 days. A significant peak at 1.299 days, corresponding to its surface rotation, was also detected in the periodogram \citep{Oliveira2022}. The difference between these periods demonstrated the existence of differential rotation in PG\,1159--035. Furthermore, a pulsation mode at 516 s in PG\,1159--035 was detected to change with a rate of $dP/dt=(-1.2\,\pm 0.1)\times 10^{-11} $s s$^{-1}$ \citep{Winget1985}. \cite{Oliveira2022} observed complex and unclear patterns in the rate of period change over the years, noting cycles of both negative and positive rate values in the $m$ components of several modes. This behaviour was suggested to be linked to non-linear mode coupling (see Sect.~\ref{sec:mode_coupling}).

\begin{table*}
    \caption{Summary of BlackGEM's observational parameters of PG\,1159--035.
    \label{tab:BG_obs_table}}
    \centering
    \begin{minipage}{16cm}
    \begin{tabular}{crrrccccc}
    \toprule
         Target name& N$q$ & N$u$ & N$i$ & Time base &$\rm BG_q$ &$\rm BG_u$ &$\rm BG_i$& Type\\
         &&&&(days)&(mag)&(mag)&(mag)&\\ \midrule
         PG\,1159--035& 414 & 201 & 197 & 712 &14.84&14.02&15.43&GW Vir
         \footnote{\small \citealt{Oliveira2022}}\footnote{\small \citealt{Winget1991}}
    \end{tabular}
    \end{minipage}    
\end{table*}

\begin{figure}
    \centering
    \includegraphics[width=0.9\linewidth]{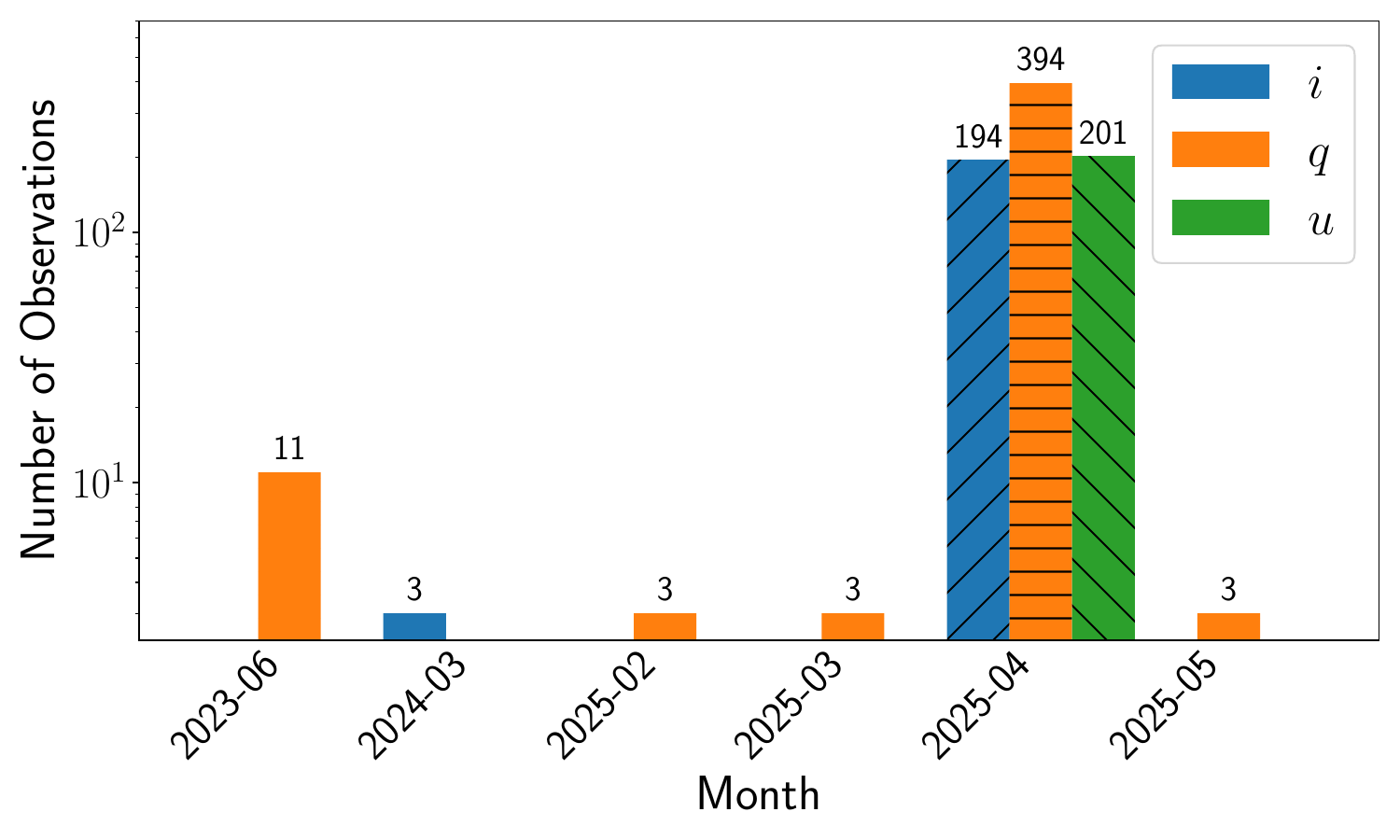}
    \caption{Number of observations per month per filter for PG\,1159--035's BlackGEM light curve.}
    \label{fig:nobs_per_month}
\end{figure}
BlackGEM observations for PG\,1159--035 were taken between June 2023 and May 2025 in the $q-, u-$, and $i-$filter bands (see \citealt{Groot2024} for the filter curve definitions), with the bulk of the data obtained during the intensive April 2025 campaign. The shorter sequences from June 2023 and May 2025 were retained to improve the frequency resolution and window function, but do not affect the main frequency content derived from the 2025 campaign. Light curves were extracted and processed using BlackGEM's forced photometry pipeline, developed by the BlackGEM software team. In brief, photometric optimal-photometry measurements, following \cite{Horne1986}, were obtained for each detection after removing nearby bright sources by iterative cleaning (Vreeswijk et al., in prep.) to reduce contamination.  
After applying light curve quality criteria to the data (see Sect. \ref{sec:prewhitening}), the cleaned light curve for PG\,1159--035 consists of 414, 201, and 197 observations in the $q, u$, and $i$ bands, respectively, as summarised in Table~\ref{tab:BG_obs_table} and Fig.~\ref{fig:nobs_per_month}. 

\begin{figure*}
    \centering
    \begin{tabular}{cc}
\includegraphics[width=0.8\linewidth]{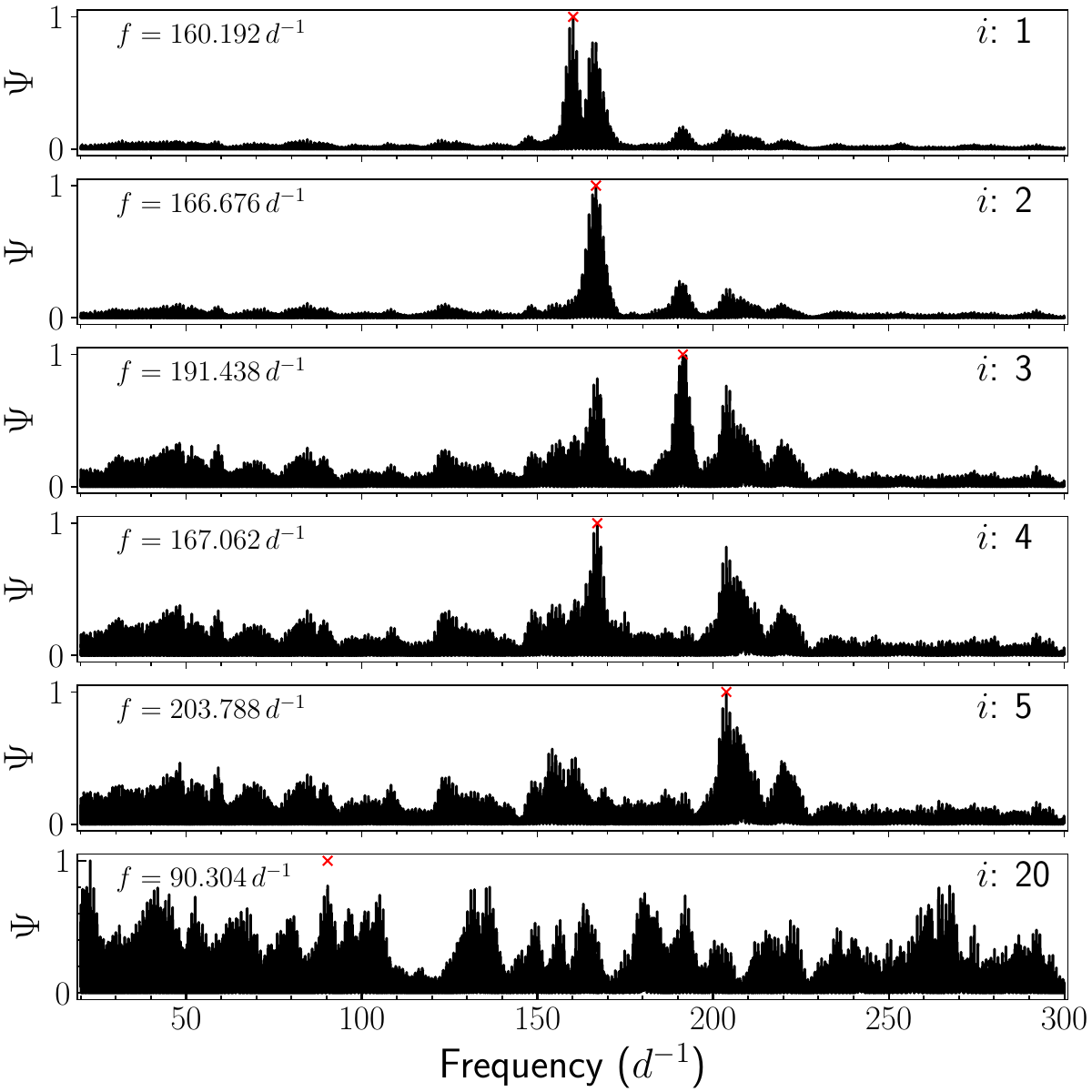}
    \end{tabular}
    \caption{ Periodograms of the first five iterations and the last iteration in the iterative pre-whitening processes of the frequencies obtained from the $q$ band.}
    \label{fig:prewhintening_top_five}
\end{figure*}

\subsection{Pulsations extractions}\label{sec:prewhitening}
A dedicated frequency analysis approach for sparsely sampled light curves has been developed and described in detail in \cite{Ranaivomanana2023} to extract the dominant frequency in hot and compact objects observed from the MeerLICHT telescope. Uncertainties on the detected frequencies were estimated using a Monte Carlo approach, as in \cite{Ranaivomanana2025}. The frequency search algorithm, known as the hybrid $\Psi$-statistic, is based on a combination of a Lomb-Scargle periodogram (LSP; \citealt{Lomb1976,Scargle1982}) and Lafler-Kinman statistic (\citealt{Clarke2002,Lafler1965}), which includes the uncertainties in the flux measurements in the computation of the periodogram. Since this technique has proven effective for MeerLICHT data, the same approach was used for the BlackGEM light curves. 

Several pre-processing steps were considered before extracting the dominant frequency in the light curves for each BlackGEM band. Observation times were converted to Barycentric Julian Date (BJD\_TDB) to correct for the light-travel time between the Earth and the Solar System barycentre \citep{Eastman2010}. This ensures a consistent barycentric time reference and minimises light-travel-time systematics, which is important when analysing high-frequency pulsations over long time baselines and when comparing frequencies with independent datasets. For the calculation of pulsational amplitudes and phases, the same initial time stamp $t_0$ was used to subtract the $q$-, $u$-, and $i$-band light curves for a consistent zero-point in time (e.g. \citealt{Beeck2021}). Additionally, observations with large photometric zero-point uncertainties were removed from the light curves. These were identified using the BlackGEM light curve quality flag, where {\tt QC-FLAG = 'red'}. Simultaneously, measurements unaffected by bad pixels or cosmic rays were selected using {\tt FLAGS\_MASK = 0}.

The frequency search was conducted from $3\,\rm d^{-1}$ to $360\,\rm d^{-1}$ with an oversampling factor of 10 using the magnitude values in each band. The lower limit of this frequency trial range was chosen to avoid low-frequency signal, as frequencies in that regime are often introduced by instrumental effects and their one-day aliases in ground-based day-night cycle observations. After extracting the dominant frequency from the processed light curve, the iterative pre-whitening was applied to extract multiple frequency peaks from the light curve using a frequency range from 20 $\rm d^{-1}$  to 300 $\rm d^{-1}$. This choice was driven by the properties of the spectral window function, which shows strong power below 20 $\rm d^{-1}$. The iterative pre-whitening steps consist of optimising the original frequency peak using increased frequency sampling (oversampling factor = 100) within a small frequency window around the peak \citep{Press1989,VanderPlas2015}. Fixing the optimised frequency, the amplitude and phase were then estimated using a least-square fit of a sinusoid model, where the uncertainties in amplitude and phase were derived using Jackknife uncertainty estimates \citep{Efron1982,Thomson2007}. The three calculated parameters (frequency, amplitude, and phase) were further optimised using a non-linear fitting algorithm based on the Levenberg-Marquardt method, implemented in the {\tt lmfit} python package \citep{Newville2016}. These three parameters were considered as free parameters in the non-linear fit, where the optimal values obtained from the least-square fit serve as the initial parameters. At each iterative process, the final model was built using all parameters, including those taken from previous iterations. 

The residual light curve was obtained by subtracting the original phase-folded light curve from the final model. Then, the pre-whitening steps above were repeated, starting from the dominant frequency search on the residual. 
The iterative pre-whitening steps were independently conducted for each band. The pre-whitening process was limited to 20 iterations to include a few low-amplitude peaks for illustrative purposes. While some of these coincide with frequencies previously reported from independent campaigns, their detection in the present dataset alone is not significant. They are retained solely to demonstrate the extraction and comparison of multi-band amplitude ratios. The derived amplitudes and phases for these marginal peaks should be regarded as uncertain, particularly given the known temporal variability of PG 1159-035’s pulsation spectrum. In this sense, several of the frequencies listed in Tables~\ref{tab:prewhitening_qband}--\ref{tab:prewhitening_uband} would not be considered significant without independent datasets. 

The reported 20 frequencies, amplitudes, and phases in Tables~\ref{tab:prewhitening_qband}--\ref{tab:prewhitening_uband} correspond to those simultaneously optimised at the end of the pre-whitening steps using the final non-linear regression model. The standard deviation of the final residual light curve in the $q$ band is 0.006 mag. 

\tabcolsep=4pt
\begin{table*}
    \centering 
    \caption{Pre-whitened frequencies obtained from the $q$-band light curve.}
    \label{tab:prewhitening_qband}
\begin{tabular}{crccrcrcrcc}
\toprule
ID              &       \multicolumn{1}{c}{$f_{\rm BG}$}        &       \multicolumn{1}{c}{$f_{\rm K2}$}   & $f_{\rm WET}$&        \multicolumn{1}{c}{Amplitude ($q$)} &   $\ell$  &\multicolumn{1}{c}{$m$} &$\ell$&\multicolumn{1}{c}{$m$} &$|f_{\rm BG}- f_{\rm K2}|$& $|f_{\rm BG}- f_{\rm WET}|$   \\      
& \multicolumn{1}{c}{(d$^{-1}$)} & \multicolumn{1}{c}{(d$^{-1}$)} &\multicolumn{1}{c}{(d$^{-1}$)} &\multicolumn{1}{c}{(mmag)} &\multicolumn{1}{c}{(K2)}&\multicolumn{1}{r}{(K2)}&\multicolumn{1}{c}{(WET)}&\multicolumn{1}{r}{(WET)}&&\\ \midrule
1       &       160.19186       $\pm$   0.00093 &       160.194 &       160.190 &       9.84    $\pm$   0.78    &       1       &       --1     &       1       &       --1     &       0.00    &       0.00    \\
2       &       166.67614       $\pm$   0.00086 &       166.686 &       166.692 &       6.63    $\pm$   0.70    &       1       &       --1     &       1       &       --1     &       0.01    &       0.02    \\
3       &       191.43774       $\pm$   0.00086 &       191.316 &       191.324 &       3.63    $\pm$   0.66    &       1       &       1       &       1       &       --1     &       0.12    &       0.11    \\
4       &       167.06157       $\pm$   0.00086 &       167.049 &       167.060 &       5.63    $\pm$   0.64    &       1       &       0       &       1       &       0       &       0.01    &       0.00    \\
5       &       203.78818       $\pm$   0.00089 &       203.886 &       203.875 &       3.39    $\pm$   0.59    &       2       &       1       &       2       &       1       &       0.10    &       0.09    \\
6       &       154.23822       $\pm$   0.00090 &       154.347 &       154.330 &       2.66    $\pm$   0.58    &       1       &       --1     &       1       &       --1     &       0.11    &       0.09    \\
7       &       207.35905       $\pm$   0.00085 &       207.895 &       207.892 &       2.31    $\pm$   0.58    &       2       &       --2     &       2       &       --2     &       0.54    &       0.53    \\
8       &       48.24380        $\pm$   0.00087 &       -       &       -       &       2.38    $\pm$   0.57    &       -       &       -       &       -       &       -       &       -       &       -       \\
9       &       159.81128       $\pm$   0.00086 &       -       &       -       &       2.80    $\pm$   0.52    &       -       &       -       &       -       &       -       &       -       &       -       \\
10      &       58.31069        $\pm$   0.00092 &       -       &       -       &       2.06    $\pm$   0.54    &       -       &       -       &       -       &       -       &       -       &       -       \\
11      &       123.86998       $\pm$   0.00089 &       -       &       -       &       2.26    $\pm$   0.52    &       -       &       -       &       -       &       -       &       -       &       -       \\
12      &       168.29738       $\pm$   0.00085 &       -       &       -       &       2.63    $\pm$   0.49    &       -       &       -       &       -       &       -       &       -       &       -       \\
13      &       222.35767       $\pm$   0.00086 &       -       &       -       &       2.15    $\pm$   0.50    &       -       &       -       &       -       &       -       &       -       &       -       \\
14      &       210.06541       $\pm$   0.00087 &       209.688 &       209.653 &       2.27    $\pm$   0.49    &       2       &       1       &       2       &       1       &       0.38    &       0.41    \\
15      &       30.95121        $\pm$   0.00088 &       -       &       -       &       2.06    $\pm$   0.49    &       -       &       -       &       -       &       -       &       -       &       -       \\
16      &       35.26341        $\pm$   0.00089 &       -       &       -       &       1.86    $\pm$   0.48    &       -       &       -       &       -       &       -       &       -       &       -       \\
17      &       186.41795       $\pm$   0.00092 &       -       &       -       &       1.84    $\pm$   0.49    &       -       &       -       &       -       &       -       &       -       &       -       \\
18      &       108.34529       $\pm$   0.00094 &       -       &       -       &       1.54    $\pm$   0.47    &       -       &       -       &       -       &       -       &       -       &       -       \\
19      &       84.53237        $\pm$   0.00091 &       -       &       -       &       1.61    $\pm$   0.43    &       -       &       -       &       -       &       -       &       -       &       -       \\
20      &       90.30412        $\pm$   0.00087 &       -       &       -       &       1.46    $\pm$   0.45    &       -       &       -       &       -       &       -       &       -       &       -       \\
\hline
\end{tabular}
\tablefoot{Column $|f_{BG}- f_{K2}|$ represents the differences in the $q$-band and K2 frequencies. Column $|f_{BG}- f_{WET}|$ corresponds to the differences in the $q$-band and WET frequencies.}
\end{table*}

\section{Results}\label{sec:result}
The extracted frequencies are presented in Tables~\ref{tab:prewhitening_qband}--\ref{tab:prewhitening_uband} for each filter band. Although the hybrid $\Psi$-statistic reduces diurnal aliasing more effectively than a Lomb–Scargle periodogram, the aliases were not entirely eliminated. Defining robust S/N-based thresholds for the $\Psi$-statistic remains non-trivial, and the low-amplitude peaks in Tables~\ref{tab:prewhitening_qband}--\ref{tab:prewhitening_uband} should not be regarded as confirmed independent frequencies without external validation. Therefore, the 20 extracted frequencies were compared with previously detected modes from K2 \citep{Oliveira2022} and WET \citep{Winget1991} observations. To ensure robust mode frequencies, only those identified in both K2 and WET were considered as a match (labelled as such in Tables~\ref{tab:prewhitening_qband}--\ref{tab:prewhitening_uband}).
The reported uncertainties in the fitted frequencies and amplitudes were estimated using Monte Carlo resampling within the iterative $\Psi$-statistic pre-whitening procedure. As this method is non-linear and incorporates irregular sampling and multi-frequency correlations, the empirical uncertainties do not strictly follow the analytical relations derived by \cite{Montgomery1999}. The reported values therefore represent the scatter obtained from repeated noise realisations, rather than formal least-squares errors. The amplitude uncertainties should be regarded as conditional estimates, derived at a fixed frequency; therefore, they represent lower limits on the full marginal uncertainties of the final multi-frequency fit. Similarly, the derived frequency uncertainties primarily reflect the common time baseline, window function, and correlated noise properties of the data, rather than the number of observed cycles alone. This naturally leads to comparable frequency uncertainties for long- and short-period modes with similar signal-to-noise ratios. For the dominant modes considered in the amplitude-ratio analysis, these uncertainties do not alter the qualitative trends discussed in this work.

To facilitate comparisons with previously published measurements, we list  the differences between the BlackGEM and K2 or WET frequencies in the two right-most columns. These differences are useful indicators of whether discrepancies may arise from one-day or half-day aliasing rather than from intrinsic frequency variability. The uncertainty associated with each difference is the quadratic sum of the individual BlackGEM and K2 or WET frequency uncertainties. For the $q$-band data, this combined uncertainty is of the order of $10^{-3}\,\rm d^{-1}$ (i.e. more than two orders of magnitude smaller than the frequency differences listed) and, thus, it does not affect the interpretation. For completeness, we also list the $i-$ and $u$-band frequency differences in Table~\ref{tab:prewhitening_iband} and Table~\ref{tab:prewhitening_uband}, noting that their uncertainties are somewhat larger owing to the lower S/N in those bands, but these measurements were not used in the subsequent analysis. 

We also note that small differences between nominally corresponding frequencies (typically several $10^{-3}\,\rm d^{-1}$) are within the frequency precision permitted by the observational cadence, window function, and the effective time baseline of the BlackGEM data. Thus, they should not be interpreted as physical differences between modes.

The frequencies from the $q$-band data that have identified modes from K2 and WET are presented in Table~\ref{tab:prewhitening_qband}, where five frequencies correspond to $\ell = 1$ and three frequencies to $\ell = 2$. These are, in most cases, consistent with those found in K2 and WET, allowing us to deduce daily aliases that are unavoidable in single-site ground-based data. In Table~\ref{tab:prewhitening_qband}, the differences between the $q$-band frequencies and the literature frequencies from K2 and WET are shown in the columns $|f_{BG}- f_{K2}|$ and $|f_{BG}- f_{WET}|$, respectively. We note that a frequency with ambiguous mode identification was not included in the list of reported modes. For instance, this is the case for the $q$-band frequency at $\rm 222.358\,d^{-1}$, which is very close (within $\rm 1\,d^{-1}$) to K2 frequencies at $\rm 223.324\,d^{-1}\,(\ell=1)$ and $\rm 221.376\,d^{-1}\,(\ell=2)$, and WET frequency at $\rm 221.368\,d^{-1} (\ell=2)$. 

Furthermore, the dominant frequencies with the highest amplitudes in K2 and WET correspond to two different frequencies, 160.902 $\rm d^{-1}$ and 167.428 $\rm d^{-1}$, respectively, with both frequencies being $\ell=1$ and $m=1$ modes. Both frequencies were not detected in the $q$-band frequencies listed in Table~\ref{tab:prewhitening_qband}. Rather, the dominant frequency in BlackGEM $q$-band data at 160.192 $\rm d^{-1}$ is  the $m=-1$ component  of the K2 dominant frequency triplet. The differences between these dominant frequencies are further discussed in Sect.~\ref{sec:mode_coupling}    

Regarding the $i$-band light curve, a total of six frequencies with only $\ell=1$ modes were identified from K2 and WET, as shown in Table~\ref{tab:prewhitening_iband}. Only three of these six modes match with those found in the $q$-band data. The dominant frequency in the $i$-band light curve is consistent with the $q$-band dominant frequency, even if it is $1\,\rm d^{-1}$ away from the $q$-band one. 

As for the $u$-band light curve, a total of five frequencies with only $\ell=1$ modes were found among those detected in K2 and WET. Three of these frequencies match the $q$-band frequencies, while two of them match the $i$-band ones.
In total, only two frequencies with $\ell=1$ modes were matched across the BlackGEM $q,\,u,$ and $i$ bands. Therefore, in the procedure described in Sect.~\ref{sec:amplitude_ratio}, the $q$-band frequencies, which have the best precision and are known modes from K2 and WET, were used to estimate the corresponding amplitudes in the $u$ and $i$ bands.

\begin{figure*}
    \centering
    \begin{tabular}{ccc}
\includegraphics[width=0.3\linewidth]{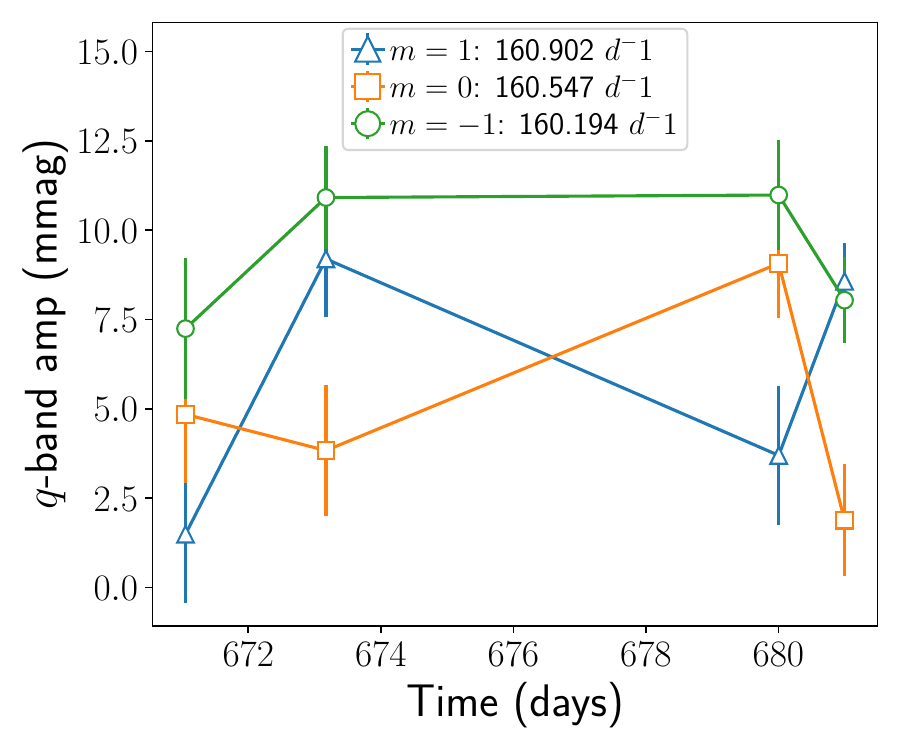}&
\includegraphics[width=0.3\linewidth]{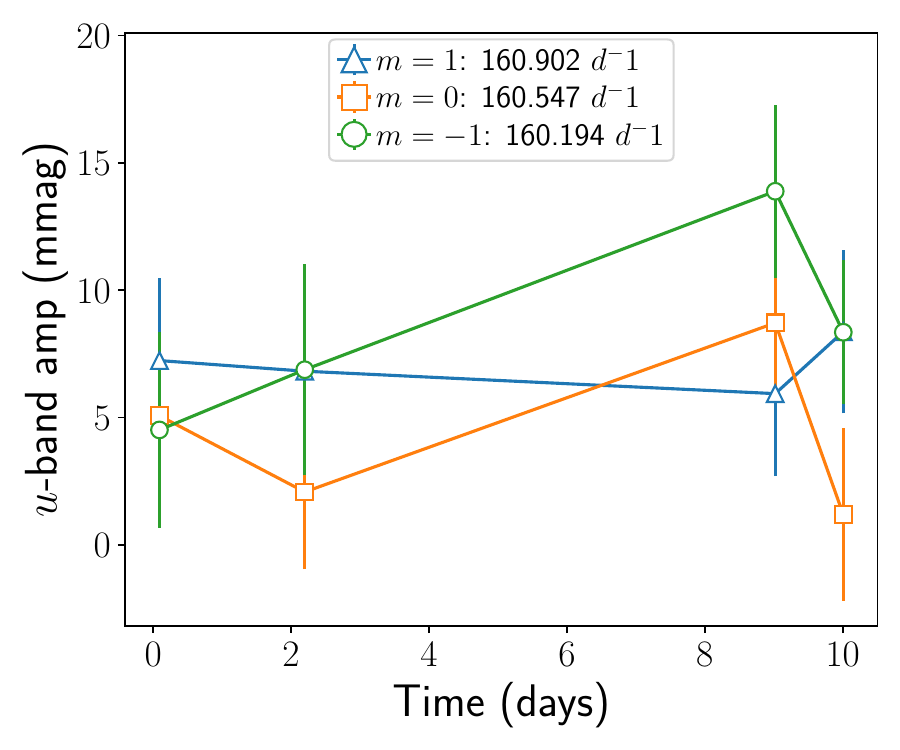}&
\includegraphics[width=0.3\linewidth]{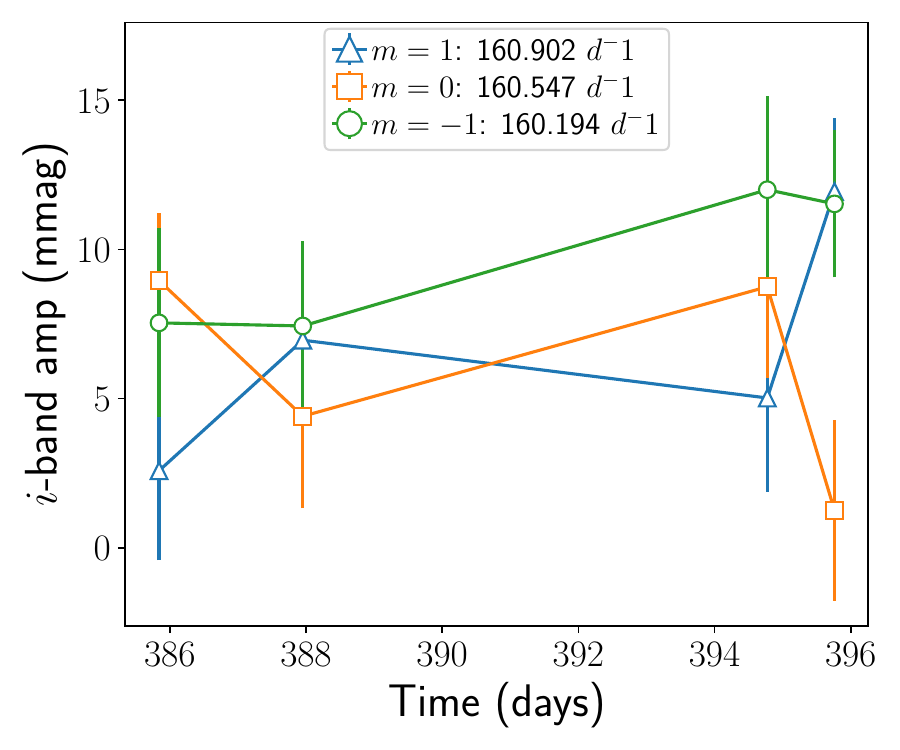}
    \end{tabular}
    \caption{Amplitude evolution of the K2 triplet mode, estimated using BlackGEM $q$-, $u$-, and $i$-band light curves. The x-axis represents the median of the time-series in each of the four time-series bins, while the y-axis their corresponding amplitudes.}
    \label{fig:resonant_mode_triplet}
\end{figure*}

\subsection{Combination frequencies and resonant mode coupling}\label{sec:mode_coupling}
The difference in the dominant frequencies across the three white-light observations (WET, K2, and TESS) and the variability within the same multiplet might indicate mode interactions and amplitude variability in PG\,1159--035. This phenomenon is often attributed to non-linear resonant mode coupling \citep{Buchler1984, Goupil1998}. Such non-linear mode interactions cause the pulsation modes to exchange energy, leading to a time-dependent increase in the amplitude of one mode at the expense of another mode's amplitude. From a theoretical point of view, such resonant mode interactions must satisfy specific resonance conditions deduced from coupled mode amplitude equations 
\citep[AEs;][]{VanHoolst1994,Mourabit2023,Mourabit2025}. For three independent frequencies, a common condition for mode resonances to occur is  $\nu_0 \sim \nu_1 + \nu_2$ \citep{Dziembowski1982,Moskalik1985}. For a rotationally split
mode triplet, the condition $\nu_+ + \nu_- \sim 2\nu_0$, with $\nu_0, \nu_-$, and $\nu_+$ corresponding to the frequencies of the $m=0, m=-1,$ and $m=1$ components, respectively, is also often 
accompanied by time-dependent amplitudes within the triplet \citep{Buchler1995, Buchler1997}.

Non-linear, coupled mode resonances have been established in numerous main-sequence gravity-mode pulsators studied from {\it Kepler\/} space photometry \citep{GangLi2020,Beeck2021}. 
On the other hand, non-linear distortion of a star's outer layers can also generate combination frequencies in the Fourier transform of the light curve \citep{Aerts2010}. In fact, such combination frequencies are nowadays omnipresent in space photometry because it is of very high precision, typically of micro-magnitude ($\micro$mag) level \citep{Kurtz2015}. However, non-linearly distorted light curves are not necessarily due to non-linear resonant mode coupling. One way to distinguish between the two is via the presence or absence of phase locking among the coupled modes and their frequencies \citep{Vuille2000,Degroote2009,Breger2014,Beeck2021}. Resonant non-linear mode coupling has already been detected in $\micro$mag-precision space photometry of both subdwarf and white dwarf pulsators \citep{Zong2016b,Zong2016a}. However, here we are dealing with millimagnitude precision, which makes a firm detection much harder to achieve.

Combination frequencies in PG\,1159--035
were first observed in K2 data by \cite{Oliveira2022}. Although the authors suggest that flux variations due to temperature perturbations \citep{Brassard1995} is one likely explanation for the combination frequencies, the reasons behind their existence have not been established thus far for PG\,1159--035. Here, we identified
two combination frequencies in the $q$ band when using a tolerance of $\Delta f<0.1 \rm d^{-1}$. We note that this tolerance was used only as a practical matching criterion and it does not imply confirmed nonlinear coupling. Given the limited frequency precision and time baseline of the present data, the identified coincidences should be regarded as tentative indications of possible frequency relationships, rather than secure detections of mode interactions.

The first combination frequency follows $f_{18} = f_{2} - f_{10}$ (following the frequency IDs in Table~\ref{tab:prewhitening_qband}), with a difference of 0.02 $\rm d^{-1}$. Here, $f_2$ is the $\ell=1$ mode frequency in Table~\ref{tab:prewhitening_qband}, while $f_{18}$ is only matched (within $\sim 0.9\,\rm d^{-1}$) with the WET frequency, 109.237 $\rm d^{-1}$, which corresponds to an $\ell=1$ mode. The third frequency, $f_{10}$, is not detected in any previous observations of PG\,1159--035 (\citealt{Winget1991,Costa2008,Oliveira2022}). 
The second combination frequency follows $f_{15}=f_{13}-f_{3}$, with a difference of $0.03\,\rm d^{-1}$. Finally, the frequency $f_3$ is an $\ell=1$ mode, $f_{13}$ is ambiguous and could be either $\ell=1$ or $\ell=2$ from K2 and WET frequencies, and $f_{15}$ is not included in either the K2 and WET frequencies.

Furthermore, the $\nu_+ +\, \nu_- \sim 2\nu_0$ relationship might explain the observed difference in the dominant components within the same triplet, where the $m=1$ and $m=-1$ components are dominant in K2 and BlackGEM observations, respectively. The K2 reported frequencies for this triplet indeed satisfy this equation: the sum of 160.194 $\rm d^{-1}$ ($\nu_-$) and 160.902 $\rm d^{-1}$ ($\nu_+$) is 321.096 $\rm d^{-1}$ and twice the central component, at 160.547 $\rm d^{-1}$, is 321.095 $\rm d^{-1}$. The difference is merely 0.00077 $\rm d^{-1}$ when considering full decimal precision.
\begin{figure}
    \centering
\includegraphics[width=0.95\linewidth]{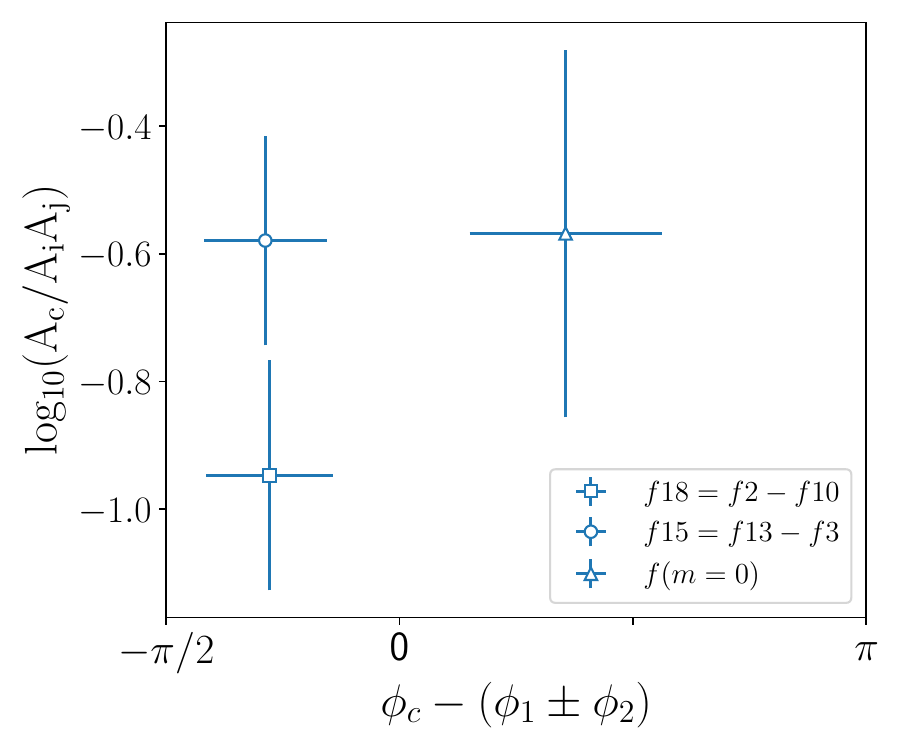}
    \caption{Relative amplitude as a function of relative phase of the two combination frequencies and the triplet mode. Parameter $f(m=0)$ represents the frequency of the central component of the triplet mode.}
    \label{fig:phase_diff_vs_relative_amp}
\end{figure}

\cite{Winget1991} and \cite{Costa2008} observed that the amplitudes of the multiplet modes change over time. To understand the origin of the two combination frequencies $f_{18}$ and $f_{15}$, along with the resonant condition in the triplet mode, evidence of amplitude modulation and phase locking was investigated to determine whether the linear combination is a result of light curve distortion or resonant mode coupling. First,
to search for signatures of amplitude modulation, the $q-, u-,$ and $i$-band light curves were divided into four time intervals, each containing approximately the same number of observations for each band. The changes of the amplitudes of the modes over time were estimated by fitting linear least-squares to these four data segments (similar to the iterative pre-whitening method), using each frequency of the triplet. The resulting amplitudes plotted against the median time of each interval are shown in Fig.~\ref{fig:resonant_mode_triplet} for the triplet mode to illustrate the amplitude variation pattern. A similar trend was found in the two combination frequencies as shown in ~\ref{fig:resonant_coupling_modes}.

The pattern of amplitude variation over time across the triplet and combination frequencies suggests evidence of amplitude modulation. The next step is to examine the amplitude and phase relations in the data. A robust way to investigate this is by analysing the relative amplitude ($A_r$) as a function of the relative phase ($\phi_r$) of the so-called parent and daughter frequencies, following \citet{Vuille2000} and \cite{Degroote2009}, and defined by $A_r=A_c/A_iA_j$ and $\phi_r=\phi_c - (n_i\phi_i+ n_j\phi_j)$, respectively. Here, the subscript $c$ represents the daughter mode, while $i$ and $j$ are the parent modes, taken to be with the largest amplitudes. In the definition of the relative phase, $\phi_r$, $n_i$, and $n_j$ correspond to the coefficients in the linear combinations \citep[see Sect.\,3.2 in ][for further details]{Degroote2009}. Combination frequencies arising from non-linear resonant mode coupling are expected to produce higher relative amplitude than those just stemming from light curve distortions. Moreover, their relative phases should be close to a multiple of $\pi/2$, including zero. The resulting values for the two combination frequencies and the triplet mode are shown in Fig.~\ref{fig:phase_diff_vs_relative_amp}. The $A_c$ values 
are quite large but with only three  measurements we cannot distinguish separate groups of daughter modes, as is the case for the large-amplitude $\beta\,$Cep pulsator HD\,180642 analysed from Convection, Rotation and planetary Transits (CoRoT; \citealt{Auvergne2009}) measurements by \cite{Degroote2009}. Given that the phase differences for the $f_{18}$ and $f_{15}$ frequency combinations do not occur at a multiple of $\pi/2$, they are likely just due to light curve distortion. 

Given the similar value for $A_c$ of the triplet and the two combination frequencies, we do not have observational evidence of any non-linear resonant mode locking at present. 
The requirements to firmly establish such non-linear resonant mode locking from amplitude and/or phase modulations are stringent in terms of the duration and precision of the light curve. While this has been established for the DB white dwarf pulsator KIC\,8626021 by \citet{Zong2016a} and for the subdwarf pulsator KIC\,10139564 \citep{Zong2016b} from four-year-long uninterrupted {\it Kepler\/} light curves, it is not (yet) within reach of the current BlackGEM data.

\begin{figure*}
    \centering
   \begin{tabular}{c}
\includegraphics[width=0.95\linewidth]{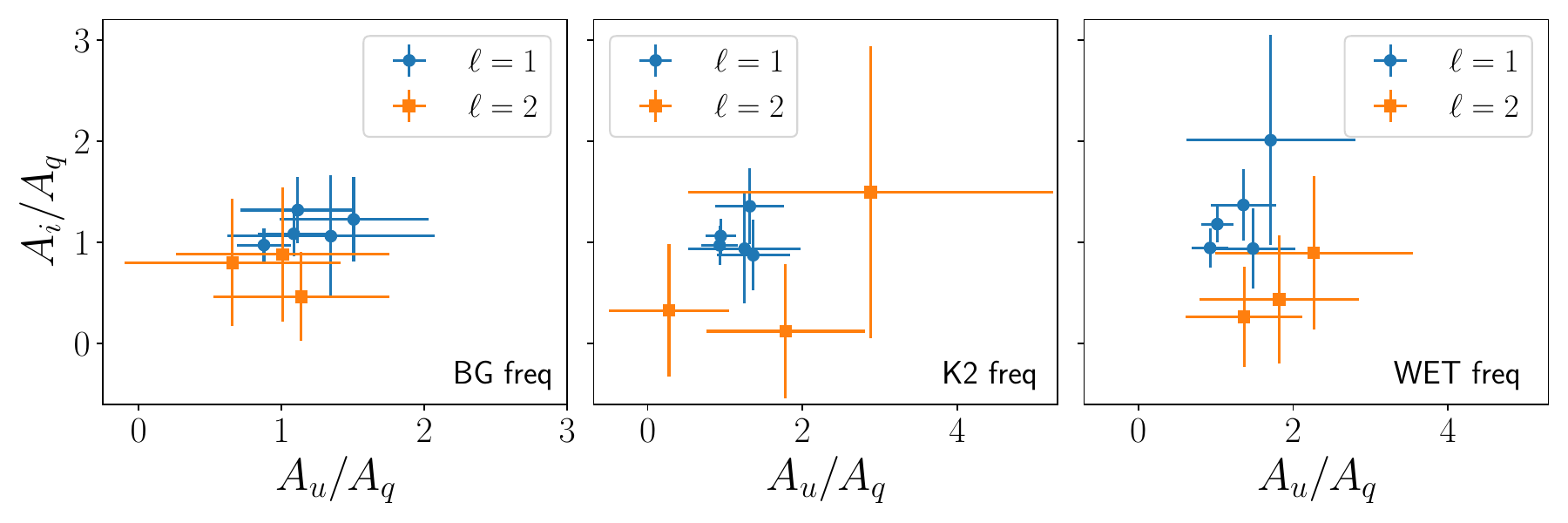}   
\end{tabular} 
    \caption{Amplitude ratio distribution of the eight pulsation modes in the $q$-band-extracted frequencies. The left panel shows the amplitude ratios derived from $q$-band frequencies, while the middle and right panels correspond to amplitude ratios estimated by imposing the K2 and WET frequencies, respectively.}
    \label{fig:amplitude_ratio}
\end{figure*}

\subsection{Amplitude ratios}\label{sec:amplitude_ratio}
To obtain amplitude ratios for the identified modes across the BlackGEM bands, it is necessary to estimate the amplitudes of the mode frequencies found in each band. Since only two frequencies were found to match in the $q,\,u,$ and $i$ bands, the eight frequencies obtained from the $q$ band were imposed on the $u-$ and $i$-band light curves to estimate their amplitudes. This approach was chosen because the $q$-band data provide the highest quality light curves in terms of cadence rate and S/N, as well as the band with the most identified mode frequencies with $\ell=1$ and $\ell=2$. The amplitude estimates were done for the $u-$ and $i-$ bands by fitting a multi-component harmonic regression model using linear least squares, considering all $q$-band frequencies in the fit. The resulting amplitude ratios for the frequencies with known modes are shown in the left panel of Fig.~\ref{fig:amplitude_ratio}, where $A_i/A_q$ is plotted against $A_u/A_q$.

To check for consistency in the distribution of the modes, the amplitudes of the $q-$, $u-$, and $i$-band light curves were estimated using the same multi-component harmonic fitting, but with imposed frequencies from the K2 and WET observations. The amplitude ratios estimated from these frequencies are shown in the middle and right panels of 
Fig.~\ref{fig:amplitude_ratio} for K2 and WET frequencies, respectively. 

The distribution of amplitude ratios derived from the $q$-band frequencies displays a trend that broadly separates the $\ell = 1$ and $\ell = 2$ modes, although some overlap remains given their respective uncertainties. This behaviour is qualitatively similar to the trends reported for main-sequence $\beta$ Cep pulsators, where \citet{Fritzewski2025} used similar data and kernel-density plots to identify modes. We note, however, that kernel-density plots allow firm mode identification only when a sufficiently large number of modes is available, as in the population study of \citet{Fritzewski2025}, which included the dominant mode of more than 100 pulsators.
In our case, we only have a few modes, but we do see that the trend observed in the left panel of Fig.~\ref{fig:amplitude_ratio} is qualitatively consistent with theirs. A comparable distribution is visible in the right panel of Fig.~\ref{fig:amplitude_ratio}, where WET frequencies were used to estimate the amplitudes.

\begin{figure*}
    \centering
\begin{tabular}{cc}
\includegraphics[width=0.45\linewidth]{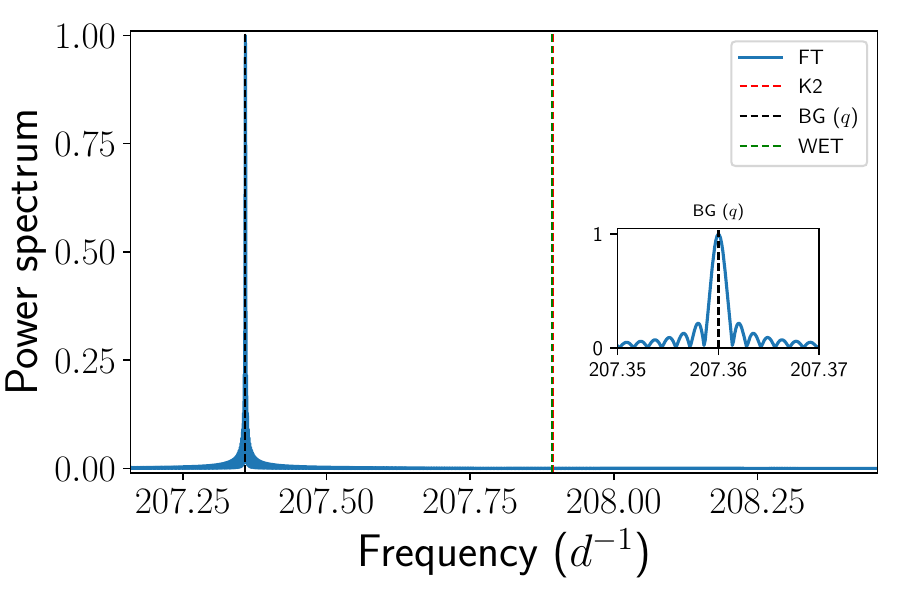} &
\includegraphics[width=0.45\linewidth]{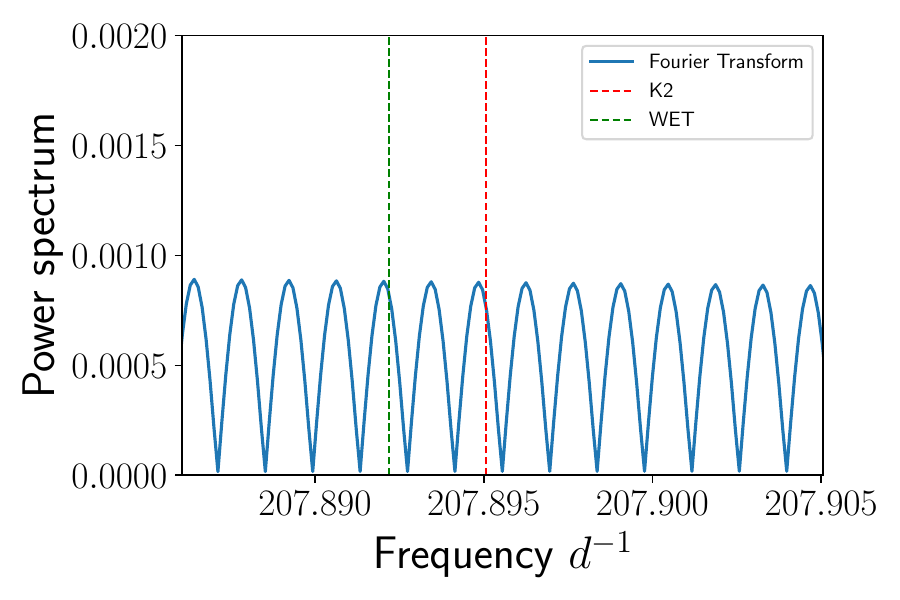} 
\end{tabular}
    \caption{Locations of the $q$-band, K2, and WET $\ell=2$ frequencies in the Fourier spectrum of the simulated sinusoidal signal using the $q$-band frequency of 207.359~$\rm d^{-1}$ (left panel). A zoomed in view of the WET and K2 frequencies is shown in the right panel.}
    \label{fig:FT}
\end{figure*}

The distribution of amplitude ratios derived from the $q$-band frequencies displays a trend that broadly separates the $\ell = 1$ and $\ell = 2$ modes, although some overlap remains given their respective uncertainties. This behaviour is qualitatively similar to the trends reported for main-sequence $\beta$ Cep pulsators, where \citet{Fritzewski2025} used similar data and kernel-density plots to identify modes. We note, however, that kernel-density plots allow firm mode identifications only when a sufficiently large number of modes is available, as in the population study of \citet{Fritzewski2025}, which included the dominant mode of more than 100 pulsators. In our case, we only have a few modes, but the trend observed in the left panel of Fig.~\ref{fig:amplitude_ratio} is qualitatively consistent with theirs. A comparable distribution is visible in the right panel of Fig.~\ref{fig:amplitude_ratio}, where WET frequencies were used to estimate the amplitudes. On the other hand, using the K2 frequencies for the amplitude ratios yields a less distinct distribution of the modes. As indicated in Table~\ref{tab:prewhitening_qband}, the K2 frequency (207.895 $\rm d^{-1}$) for this particular mode is slightly off from the WET frequency (207.892 $\rm d^{-1}$), and is about a half per day off from the $q$-band frequency (207.359 $\rm d^{-1}$), which could explain its position in the amplitude ratio distribution. This is further confirmed by computing the Fourier transform of a simulated sinusoidal signal using the $q$-band frequency. Figure~\ref{fig:FT} illustrates the position of the K2 and WET frequencies in the side lobes of the Fourier spectrum for the BlackGEM $q$ band, showing that the K2 frequency is slightly further from the centre of the side lobe than the WET frequency for this mode. 

Regarding the residuals of the fits, no significant differences were found between the observations with regards to the root mean square (RMS) of the residuals of the multi-component harmonic models, as shown in Table~\ref{tab:rms_residual}. Overall, the $q$-band light curves produced the best fit compared to the $u$ and $i$ bands, regardless of the observational frequencies used.

\begin{table*}
    \centering
    \caption{Pre-whitened frequencies obtained from the $i$-band light curve.}
    \label{tab:prewhitening_iband}
    \begin{tabular}{crrrrcrcrcc}
\toprule
ID              &       \multicolumn{1}{c}{$f_{\rm BG}$}        &       \multicolumn{1}{c}{$f_{\rm K2}$}   & $f_{\rm WET}$&        \multicolumn{1}{c}{Amplitude ($i$)} &   $\ell$  &\multicolumn{1}{c}{$m$} &$\ell$&\multicolumn{1}{c}{$m$} &$|f_{\rm BG}- f_{\rm K2}|$& $|f_{\rm BG}- f_{\rm WET}|$   \\      
& \multicolumn{1}{c}{(d$^{-1}$)} & \multicolumn{1}{c}{(d$^{-1}$)} &\multicolumn{1}{c}{(d$^{-1}$)} &\multicolumn{1}{c}{(mmag)} &\multicolumn{1}{c}{(K2)}&\multicolumn{1}{r}{(K2)}&\multicolumn{1}{c}{(WET)}&\multicolumn{1}{r}{(WET)}&&\\ \midrule
1               &       159.1975        $\pm$   0.0017  &       160.194 &       160.190 &       10.51   $\pm$   1.37    &       1       &       --1     &       1       &       --1     &       1.00    &       0.99    \\
2               &       167.8120        $\pm$   0.0017  &       -       &       -       &       6.65    $\pm$   1.22    &       -       &       -       &       -       &       -       &       -       &       -       \\
3               &       191.3361        $\pm$   0.0017  &       191.316 &       191.324 &       5.39    $\pm$   1.10    &       1       &       1       &       1       &       --1     &       0.02    &       0.01    \\
4               &       161.8975        $\pm$   0.0017  &       160.902 &       160.927 &       6.84    $\pm$   1.10    &       1       &       1       &       1       &       1       &       1.00    &       0.97    \\
5               &       226.3028        $\pm$   0.0017  &       -       &       -       &       4.35    $\pm$   1.02    &       -       &       -       &       -       &       -       &       -       &       -       \\
6               &       43.8550 $\pm$   0.0018  &       -       &       -       &       4.40    $\pm$   1.00    &       -       &       -       &       -       &       -       &       -       &       -       \\
7               &       153.4437        $\pm$   0.0017  &       154.347 &       154.330 &       2.96    $\pm$   1.02    &       1       &       --1     &       1       &       --1     &       0.90    &       0.89    \\
8               &       66.0962 $\pm$   0.0017  &       -       &       -       &       3.43    $\pm$   1.02    &       -       &       -       &       -       &       -       &       -       &       -       \\
9               &       58.4796 $\pm$   0.0018  &       -       &       -       &       2.63    $\pm$   1.00    &       -       &       -       &       -       &       -       &       -       &       -       \\
10              &       98.2781 $\pm$   0.0018  &       -       &       -       &       5.41    $\pm$   0.87    &       -       &       -       &       -       &       -       &       -       &       -       \\
11              &       146.9469        $\pm$   0.0017  &       -       &       -       &       2.78    $\pm$   0.87    &       -       &       -       &       -       &       -       &       -       &       -       \\
12              &       190.9084        $\pm$   0.0017  &       190.962 &       190.969 &       3.17    $\pm$   0.86    &       1       &       0       &       1       &       -       &       0.05    &       0.06    \\
13              &       132.8476        $\pm$   0.0017  &       -       &       -       &       2.68    $\pm$   0.98    &       -       &       -       &       -       &       -       &       -       &       -       \\
14              &       20.0222 $\pm$   0.0017  &       -       &       -       &       2.84    $\pm$   0.70    &       -       &       -       &       -       &       -       &       -       &       -       \\
15              &       228.5388        $\pm$   0.0017  &       -       &       -       &       3.23    $\pm$   0.76    &       -       &       -       &       -       &       -       &       -       &       -       \\
16              &       20.4329 $\pm$   0.0018  &       -       &       -       &       2.24    $\pm$   0.77    &       -       &       -       &       -       &       -       &       -       &       -       \\
17              &       54.7781 $\pm$   0.0017  &       -       &       -       &       2.45    $\pm$   0.75    &       -       &       -       &       -       &       -       &       -       &       -       \\
18              &       293.8998        $\pm$   0.0017  &       -       &       -       &       2.12    $\pm$   0.74    &       -       &       -       &       -       &       -       &       -       &       -       \\
19              &       87.5983 $\pm$   0.0017  &       87.407  &       87.420  &       1.75    $\pm$   0.72    &       1       &       --1     &       1       &       --1     &       0.19    &       0.18    \\
20              &       76.9228 $\pm$   0.0017  &       -       &       -       &       2.62    $\pm$   0.71    &       -       &       -       &       -       &       -       &       -       &       -       \\
\hline
\end{tabular}\tablefoot{The column $|f_{\rm BG}- f_{\rm K2}|$ represents the differences in the $i$-band and K2 frequencies, while the column $|f_{\rm BG}- f_{\rm WET}|$ corresponds to the differences in the $i$-band and WET frequencies.}
\end{table*}

\begin{table*}
    \centering
        \caption{Pre-whitened frequencies obtained from the $u$-band light curve.}
    \label{tab:prewhitening_uband}
    \begin{tabular}{crccrcrcrcc}
\toprule
ID              &       \multicolumn{1}{c}{$f_{\rm BG}$}        &       \multicolumn{1}{c}{$f_{\rm K2}$}   & $f_{\rm WET}$&        \multicolumn{1}{c}{Amplitude ($u$)} &   $\ell$  &\multicolumn{1}{c}{$m$} &$\ell$&\multicolumn{1}{c}{$m$} &$|f_{\rm BG}- f_{\rm K2}|$& $|f_{\rm BG}- f_{\rm WET}|$\\ 
& \multicolumn{1}{c}{(d$^{-1}$)} & \multicolumn{1}{c}{(d$^{-1}$)} &\multicolumn{1}{c}{(d$^{-1}$)} &\multicolumn{1}{c}{(mmag)} &\multicolumn{1}{c}{(K2)}&\multicolumn{1}{r}{(K2)}&\multicolumn{1}{c}{(WET)}&\multicolumn{1}{r}{(WET)}&&\\ \midrule
1       &       164.815 $\pm$   0.065   &       166.686 &       166.692 &       8.46    $\pm$   1.70    &       1       &       $-$1    &       1       &       $-$1    &       1.87    &       1.88    \\
2       &       160.087 $\pm$   0.062   &       160.194 &       160.190 &       8.53    $\pm$   1.52    &       1       &       $-$1    &       1       &       $-$1    &       0.11    &       0.10    \\
3       &       190.440 $\pm$   0.066   &       190.962 &       190.350 &       7.23    $\pm$   1.51    &       1       &       1       &       1       &       -       &       0.52    &       0.09    \\
4       &       92.274  $\pm$   0.064   &       -       &       -       &       7.33    $\pm$   1.44    &       -       &       -       &       -       &       -       &       -       &       -       \\
5       &       59.066  $\pm$   0.063   &       -       &       -       &       4.50    $\pm$   1.40    &       -       &       -       &       -       &       -       &       -       &       -       \\
6       &       43.755  $\pm$   0.064   &       -       &       -       &       4.21    $\pm$   1.39    &       -       &       -       &       -       &       -       &       -       &       -       \\
7       &       22.573  $\pm$   0.065   &       -       &       -       &       5.54    $\pm$   1.34    &       -       &       -       &       -       &       -       &       -       &       -       \\
8       &       133.204 $\pm$   0.064   &       134.321 &       134.285 &       4.34    $\pm$   1.62    &       1       &       0       &       1       &       0       &       1.12    &       1.08    \\
9       &       222.740 $\pm$   0.067   &       -       &       -       &       3.96    $\pm$   1.32    &       -       &       -       &       -       &       -       &       -       &       -       \\
10      &       112.845 $\pm$   0.063   &       -       &       -       &       3.80    $\pm$   1.23    &       -       &       -       &       -       &       -       &       -       &       -       \\
11      &       199.907 $\pm$   0.064   &       -       &       -       &       3.70    $\pm$   1.20    &       -       &       -       &       -       &       -       &       -       &       -       \\
12      &       298.572 $\pm$   0.064   &       -       &       -       &       3.19    $\pm$   1.25    &       -       &       -       &       -       &       -       &       -       &       -       \\
13      &       167.440 $\pm$   0.065   &       167.408 &       167.429 &       4.35    $\pm$   1.11    &       1       &       1       &       1       &       1       &       0.03    &       0.01    \\
14      &       88.044  $\pm$   0.064   &       -       &       -       &       2.75    $\pm$   1.10    &       -       &       -       &       -       &       -       &       -       &       -       \\
15      &       71.874  $\pm$   0.064   &       -       &       -       &       4.09    $\pm$   1.07    &       -       &       -       &       -       &       -       &       -       &       -       \\
16      &       84.327  $\pm$   0.064   &       -       &       -       &       3.77    $\pm$   1.05    &       -       &       -       &       -       &       -       &       -       &       -       \\
17      &       55.533  $\pm$   0.064   &       -       &       -       &       3.18    $\pm$   1.05    &       -       &       -       &       -       &       -       &       -       &       -       \\
18      &       158.015 $\pm$   0.063   &       -       &       -       &       3.11    $\pm$   1.10    &       -       &       -       &       -       &       -       &       -       &       -       \\
19      &       20.012  $\pm$   0.065   &       -       &       -       &       3.32    $\pm$   0.99    &       -       &       -       &       -       &       -       &       -       &       -       \\
20      &       23.977  $\pm$   0.064   &       -       &       -       &       3.19    $\pm$   0.99    &       -       &       -       &       -       &       -       &       -       &       -       \\
\hline
    \end{tabular}
    \tablefoot{The column $|f_{BG}- f_{K2}|$ represents the differences in the $u$-band and K2 frequencies, while the column $|f_{BG}- f_{WET}|$ corresponds to the differences in the $u$-band and WET frequencies.}
\end{table*}

\begin{table}[htbp]
\caption{Root-mean-square (RMS) residuals.}
    \label{tab:rms_residual}
    \centering
    \begin{tabular}{cccc}
         Dataset& rms$_q$ & rms$_u$ & rms$_i$\\\toprule
         BG ($q$)& - & 0.0130 & 0.0103\\
         K2& 0.0089 & 0.0137 &0.0115 \\
         WET& 0.0093 & 0.0138 & 0.0117\\\hline
    \end{tabular}
    \tablefoot{RMS of the residuals (in mag) from the multi-component harmonic regression model fit to BlackGEM light curves using $q$-band, K2, and WET pulsation mode frequencies.}
\end{table}

\section{Conclusion and future prospects}\label{sec:conclusion_chp5}
The main goal of the current study is to assess BlackGEM's capability of detecting multi-periodic light variations at millimag levels and to demonstrate the feasibility of using its multi-colour photometry to identify pulsation modes. By choosing a target with known pulsation modes from the literature for the BlackGEM observations, we have been able to compare the extracted frequencies from BlackGEM light curves with those reported previously. Our selected prototype validation target, PG\,1159--035, was specifically chosen to investigate BlackGEM's  power of amplitude ratios for mode identification. 

In this pilot application, we first extracted the frequencies from the $q$ band, providing the highest photometric precision, and then imposed these $q$-band frequencies on the $u-$ and $i$-band light curves to derive the corresponding amplitudes and phases. This procedure ensured that the same physical frequencies were used across filters, preventing spurious detections in the noisier $u$ and $i$ bands. We emphasise that the low-amplitude $u-$ and $i$-band measurements do not constitute independent frequency detections, but they were used to test the multi-colour amplitude-ratio method for this single target. In future applications to larger and higher-S/N BlackGEM samples, the amplitude-ratio analysis will be restricted to the strongest and most confidently detected modes in each band.

Frequency extractions using the $q$-band light curve revealed five $\ell=1$ and $\ell=2$ pulsation modes, as confirmed by K2 and WET reported frequencies. Six $\ell=1$ modes were detected in the $i$ band, while five $\ell=1$ modes were identified in the $u$-band light curves.

Additionally, two new combination frequencies were identified in PG\,1159--035 using the $q$-band frequencies, supplementing those obtained in a previous study \citep{Oliveira2022}. Our BlackGEM data did not allow us to conclude on 
non-linear resonant mode coupling as the origin of the combination frequencies
as no firm phase locking could not be established. This requires long, uninterrupted photometric data with high enough precision to properly evaluate the amplitudes and phase modulations of daughter modes caused by resonantly coupled parent modes. Long-term follow-up photometry with BlackGEM might shed more light on this matter.

Amplitude ratios, $A_i/A_q$ and $A_u/A_q$, were obtained using frequencies from BlackGEM, K2, and WET. This resulted in qualitatively different density distributions of the $\ell=1$ and $\ell=2$ modes identified in the $q$-band data; however, given the uncertainties on the amplitude ratios, this distinction should be regarded as illustrative rather than statistically significant. This suggests that despite the one- or half-day aliases on the BlackGEM frequencies, mode identification using amplitude ratios is potentially possible and robust against these aliases. Additionally, the frequencies extracted from the $q$-band-produced residuals comparable to those from the K2 and WET frequencies when a multi-component linear regression was used to estimate amplitudes for the $u-$ and $i$-band light curves. This confirms the potential of the $q$-band photometry for future asteroseismology of compact pulsators. The consistency observed in the residuals also demonstrates BlackGEM's capability for multi-colour variability analysis, which is essential for mode identification using amplitude ratios. This further implies that frequencies determined in the $q$-band are sufficient to derive amplitudes in other bands, eliminating the need for individual frequency searches in each band. In line with the discussion above, this approach is applicable primarily to dominant modes with amplitudes well above the noise level.

This proof-of-concept study demonstrates the power of multi-colour BlackGEM photometry in characterising and identifying pulsation modes in non-radial pulsating stars such as PG\,1159--035. Hence, BlackGEM observations may significantly contribute to the identification of these compact pulsators and allow for future powerful exploitation of observational amplitude ratios. Even with the photometric data used here coming from standard survey products (and not dedicated to asteroseismology), the findings presented in this work are promising for asteroseismic studies of these stars, opening up the faint sky to asteroseismology. As the BlackGEM Fast Synoptic Survey is projected to cover approximately 4000 square degrees of sky (2.7 square degrees per night) over its five-year period, a large number of compact pulsating objects are expected to be observed at this level of photometric quality. This includes many known and yet-to-be-identified pulsating sdB stars in the southern sky ($\delta < 30^\circ$), which are well within BlackGEM’s magnitude range. The present study  paves the way for large-scale asteroseismic analyses of such objects, ultimately contributing to our understanding of their internal structures and evolution.

\begin{acknowledgements}
Based on observations with the BlackGEM telescope array. The BlackGEM telescope array is built and run by a consortium consisting of Radboud University, the Netherlands Research School for Astronomy (NOVA), and KU Leuven with additional support from Armagh Observatory and Planetarium, Durham University, Hamburg Observatory, Hebrew University, Las Cumbres Observatory, Tel Aviv University, Texas Tech University, Technical University of Denmark, University of California Davis, the University of Barcelona, the University of Manchester, University of Potsdam, the University of Valparaiso, the University of Warwick, and Weizmann Institute of Science. BlackGEM is hosted and supported by the European Southern Observatory at La Silla, Chile.
C.J.~acknowledges funding from the Royal Society through the Newton International Fellowship funding scheme (project No. NIF$\backslash$R1$\backslash$242552). This research was supported by Deutsche Forschungsgemeinschaft  (DFG, German Research Foundation) under Germany’s Excellence Strategy - EXC 2121 "Quantum Universe" - 390833306. Co-funded by the European Union (ERC, CompactBINARIES, 101078773). Views and opinions expressed are however those of the author(s) only and do not necessarily reflect those of the European Union or the European Research Council. Neither the European Union nor the granting authority can be held responsible for them. The research leading to these results has received funding from the Research Foundation Flanders (FWO) under grant agreement
G0A2917N (BlackGEM), as well as from the BELgian federal Science Policy Office (BELSPO) through PRODEX grants for {\it Gaia} data exploitation. CA and CJ acknowledge support from Long term structural funding - Methusalem funding by the Flemish Government, project SOUL: Stellar evolution in full glory, grant METH/24/012, at KU Leuven, Belgium. This work has made use of data from the European Space Agency (ESA) mission {\it Gaia} (https://www.cosmos.esa.int/Gaia), processed by the {\it Gaia} Data Processing and Analysis Consortium (DPAC, https://www.cosmos.esa.int/web/Gaia/dpac/consortium). Funding for the DPAC has been provided by national institutions, in particular the institutions participating in the {\it Gaia} Multilateral Agreement. PJG is supported by NRF SARChI grant 111692. 

\end{acknowledgements}
\bibliographystyle{aa}
\bibliography{main}

\begin{appendix}
\onecolumn
\section{Additional material}\label{appendix:A_chp5}
\begin{figure}[hb]
    \centering
    \includegraphics[width=0.95\linewidth]{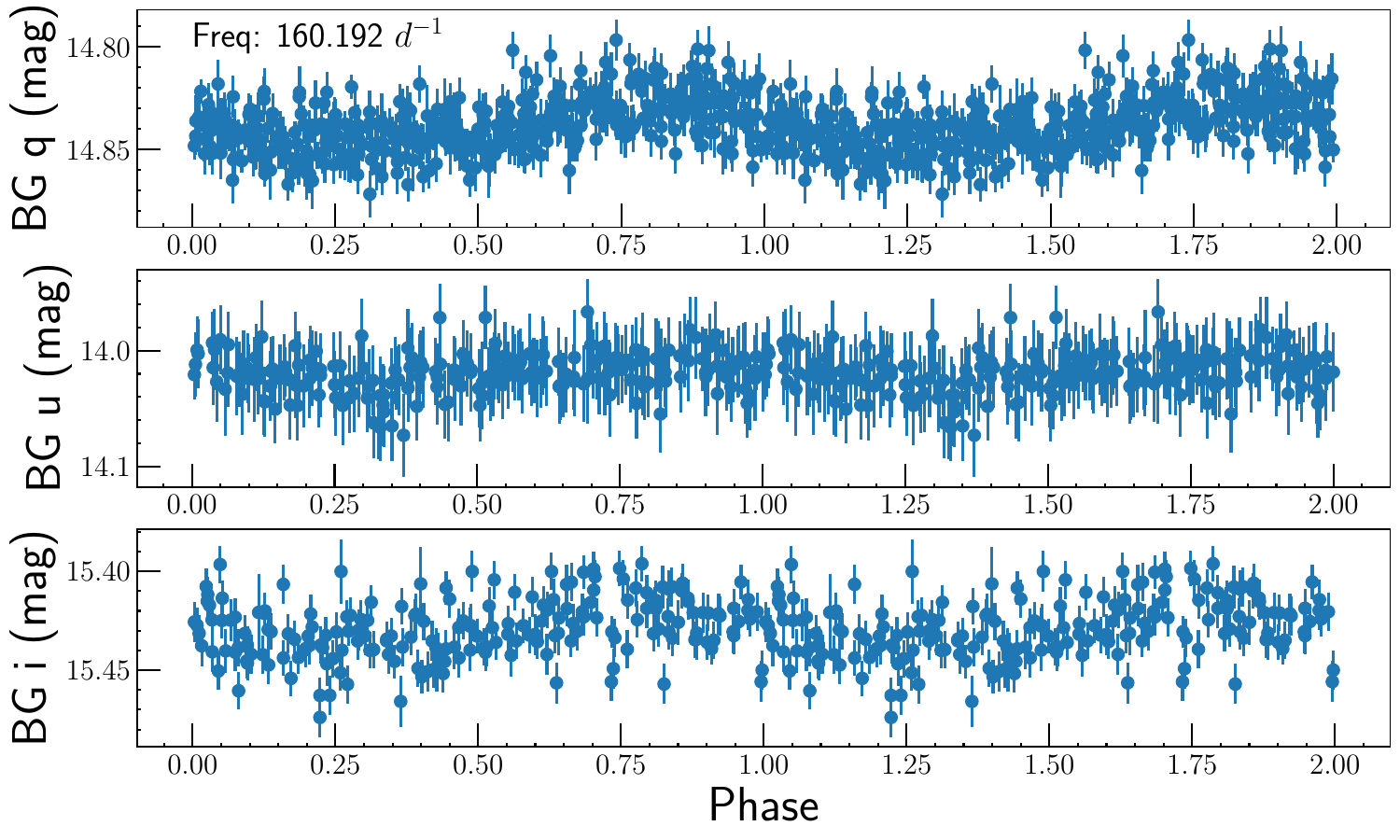}
    \caption{BlackGEM light curves of PG\,1159--035, folded at the dominant $q$-band frequency.}
    \label{fig:pg1159_BG_lc}
\end{figure}

\begin{figure}[hb]
    \centering
    \begin{tabular}{ccc}
      \includegraphics[width=0.32\linewidth]{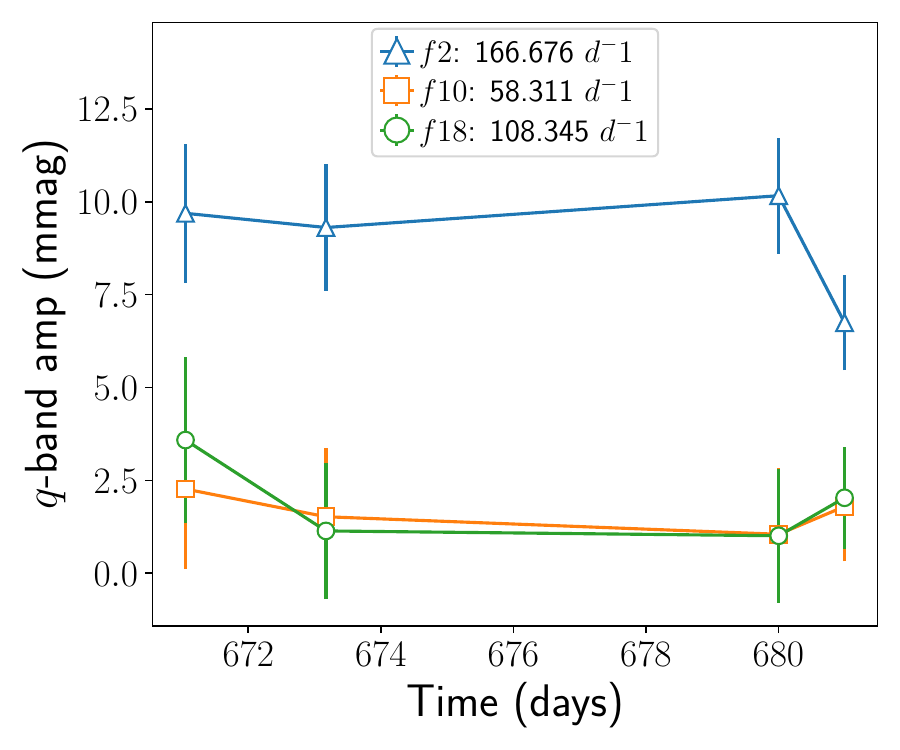}       &        \includegraphics[width=0.32\linewidth]{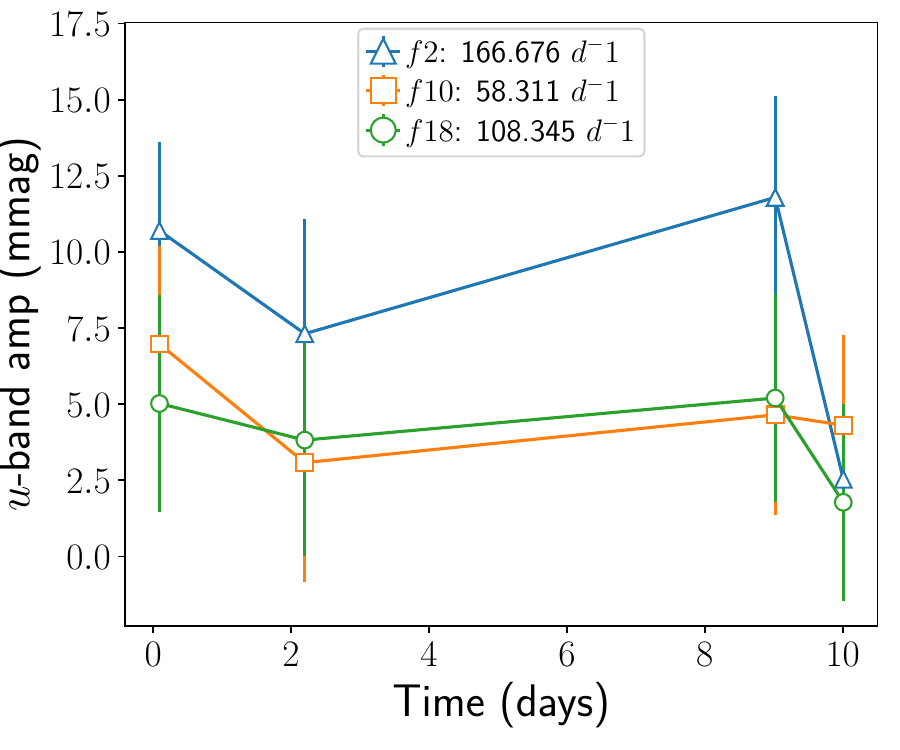}  &       \includegraphics[width=0.32\linewidth]{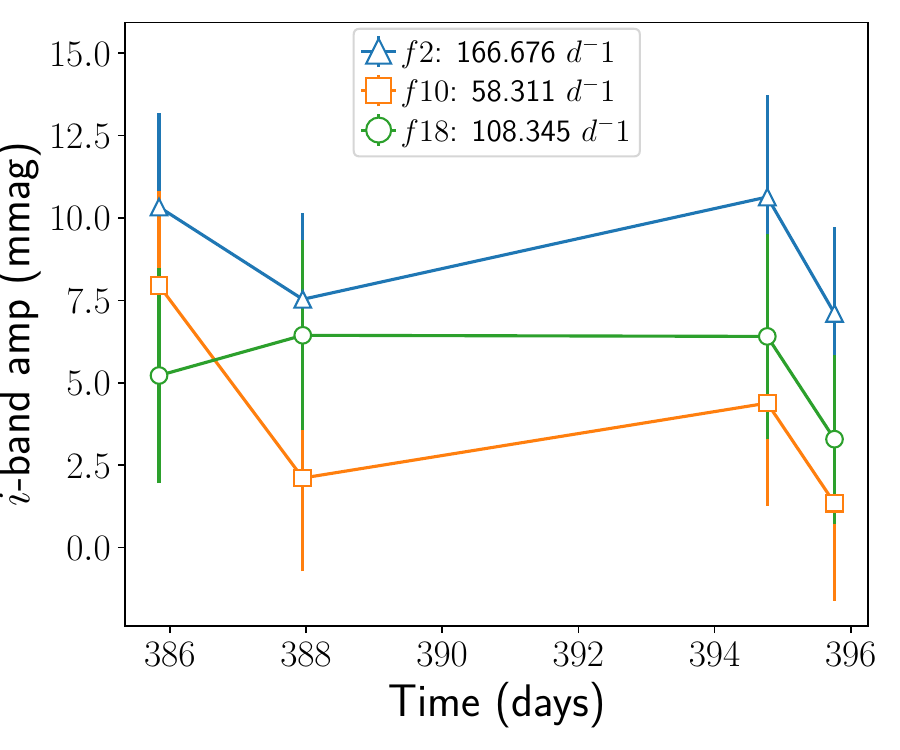}\\
    \includegraphics[width=0.32\linewidth]{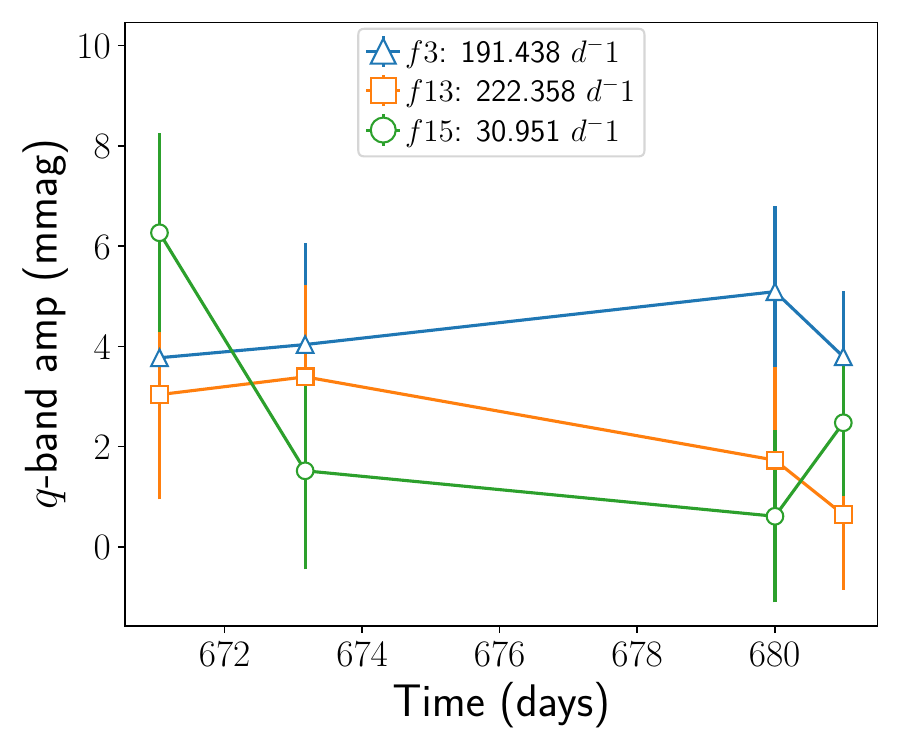} &     \includegraphics[width=0.32\linewidth]{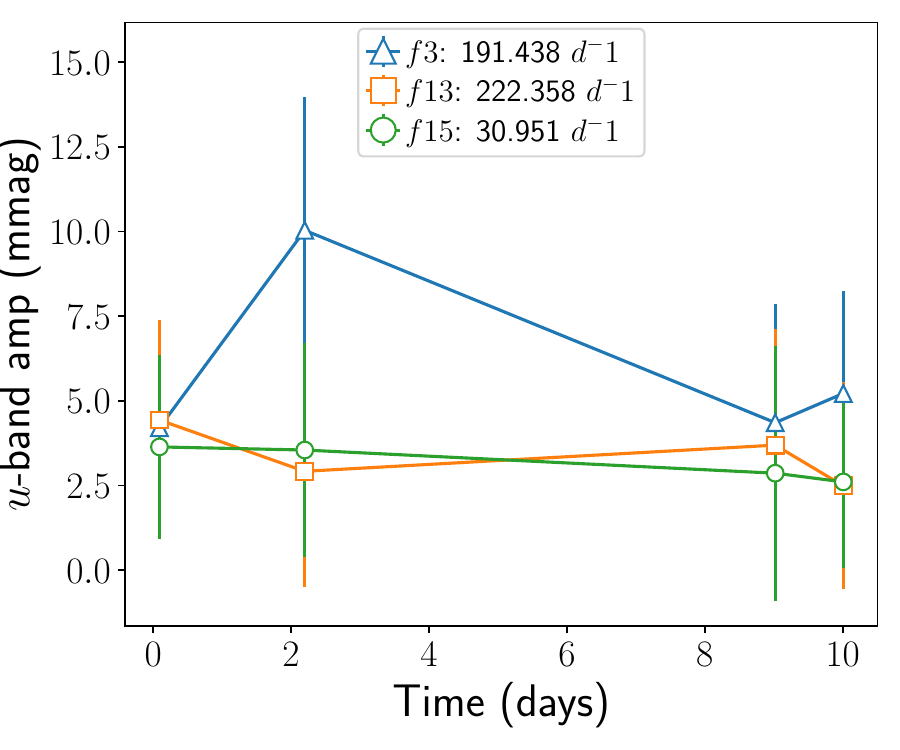}  &     \includegraphics[width=0.32\linewidth]{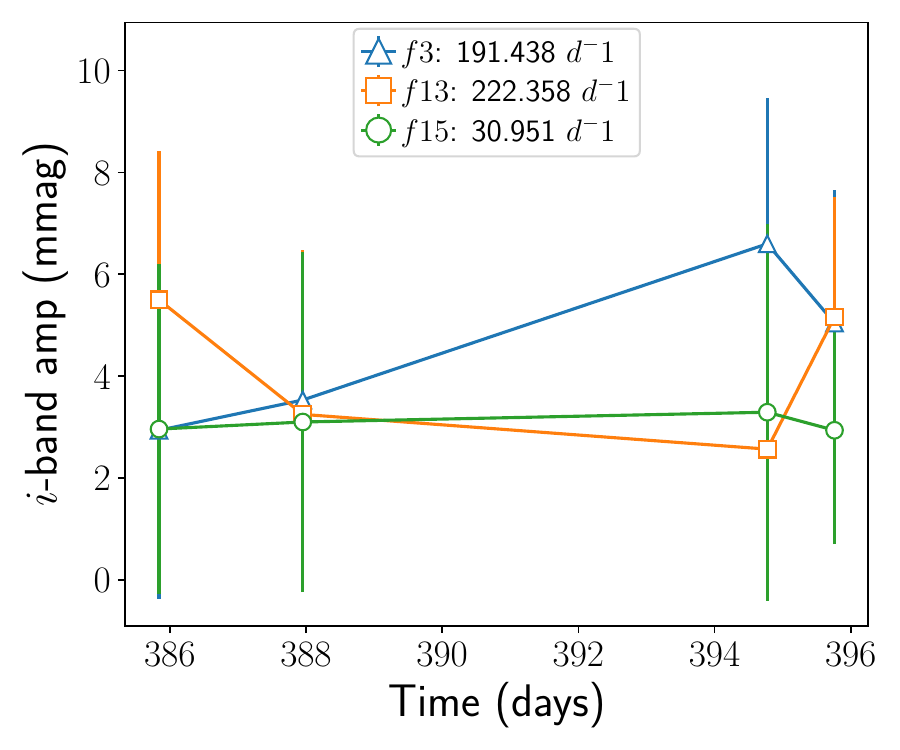} 
    \end{tabular}
    \caption{Amplitude evolution over time for the two combination frequencies identified here.}
    \label{fig:resonant_coupling_modes}
\end{figure}

\begin{table*}
\caption{Estimated amplitudes in the $u$ and $i$ bands and their ratios.}
 \resizebox{\textwidth}{!}{
    \begin{tabular}{crccccccc}
    \toprule
        ID      &       Freq. ($q$)     &       $A_q$ & $A_u$&  $A_i$&  $A_\mathrm{K2}$&        $A{_u}/A{_q}$           &       $A_{i}/A_{q}$   &       $\ell$  \\      
    & ($\rm d^{-1}$) & (mmag) &(mmag) & (mmag) & (mmag) & & & \\ \midrule
1       &       160.19186       $\pm$   0.00093 &       9.84    $\pm$   0.78    &       8.63    $\pm$   1.74    &       9.58    $\pm$   1.43    &       4.74    $\pm$   0.02    &       0.88    $\pm$   0.19    &       0.97    $\pm$   0.16    &       1       \\
2       &       166.67614       $\pm$   0.00086 &       6.63    $\pm$   0.70    &       7.20    $\pm$   1.50    &       7.19    $\pm$   1.28    &       1.46    &       1.09    $\pm$   0.25    &       1.08    $\pm$   0.22    &       1       \\
3       &       191.43774       $\pm$   0.00086 &       3.63    $\pm$   0.66    &       5.46    $\pm$   1.61    &       4.45    $\pm$   1.27    &       2.70    $\pm$   0.02    &       1.51    $\pm$   0.52    &       1.23    $\pm$   0.42    &       1       \\
4       &       167.06157       $\pm$   0.00086 &       5.63    $\pm$   0.64    &       6.27    $\pm$   2.16    &       7.43    $\pm$   1.66    &       5.34    $\pm$   0.02    &       1.11    $\pm$   0.40    &       1.32    $\pm$   0.33    &       1       \\
5       &       203.78818       $\pm$   0.00089 &       3.39    $\pm$   0.59    &       3.86    $\pm$   1.97    &       1.57    $\pm$   1.46    &       0.94    $\pm$   0.02    &       1.14    $\pm$   0.61    &       0.46    $\pm$   0.44    &       2       \\
6       &       154.23822       $\pm$   0.00090 &       2.66    $\pm$   0.58    &       3.58    $\pm$   1.76    &       2.83    $\pm$   1.48    &       0.27            &       1.35    $\pm$   0.72    &       1.06    $\pm$   0.60    &       1       \\
7       &       207.35905       $\pm$   0.00085 &       2.31    $\pm$   0.58    &       2.33    $\pm$   1.62    &       2.04    $\pm$   1.44    &       0.48    $\pm$   0.02    &       1.01    $\pm$   0.75    &       0.88    $\pm$   0.66    &       2       \\
8       &       48.24380        $\pm$   0.00087 &       2.38    $\pm$   0.57    &       0.79    $\pm$   1.65    &       0.98    $\pm$   1.29    & - &     0.33    $\pm$   0.70    &       0.41    $\pm$   0.55    &       -       \\
9       &       159.81128       $\pm$   0.00086 &       2.80    $\pm$   0.52    &       0.87    $\pm$   1.62    &       4.59    $\pm$   1.40    & - &     0.31    $\pm$   0.58    &       1.64    $\pm$   0.58    &       -       \\
10      &       58.31069        $\pm$   0.00092 &       2.06    $\pm$   0.54    &       2.60    $\pm$   1.79    &       0.36    $\pm$   1.48    & - &     1.26    $\pm$   0.93    &       0.18    $\pm$   0.72    &       -       \\
11      &       123.86998       $\pm$   0.00089 &       2.26    $\pm$   0.52    &       1.77    $\pm$   1.77    &       0.83    $\pm$   1.16    & - &     0.78    $\pm$   0.80    &       0.37    $\pm$   0.52    &       -       \\
12      &       168.29738       $\pm$   0.00085 &       2.63    $\pm$   0.49    &       3.17    $\pm$   2.00    &       5.41    $\pm$   1.60    & - &     1.21    $\pm$   0.80    &       2.06    $\pm$   0.72    &       -       \\
13      &       222.35767       $\pm$   0.00086 &       2.15    $\pm$   0.50    &       3.47    $\pm$   1.47    &       3.45    $\pm$   1.30    & - &     1.61    $\pm$   0.78    &       1.60    $\pm$   0.71    &       -       \\
14      &       210.06541       $\pm$   0.00087 &       2.27    $\pm$   0.49    &       1.49    $\pm$   1.68    &       1.82    $\pm$   1.37    &       0.24            &       0.66    $\pm$   0.75    &       0.80    $\pm$   0.63    &       2       \\
15      &       30.95121        $\pm$   0.00088 &       2.06    $\pm$   0.49    &       1.38    $\pm$   1.50    &       0.90    $\pm$   1.23    & - &     0.67    $\pm$   0.75    &       0.44    $\pm$   0.61    &       -       \\
16      &       35.26341        $\pm$   0.00089 &       1.86    $\pm$   0.48    &       1.76    $\pm$   1.67    &       2.86    $\pm$   1.34    & - &     0.94    $\pm$   0.93    &       1.53    $\pm$   0.82    &       -       \\
17      &       186.41795       $\pm$   0.00092 &       1.84    $\pm$   0.49    &       1.04    $\pm$   1.64    &       1.79    $\pm$   1.21    & - &     0.57    $\pm$   0.90    &       0.97    $\pm$   0.71    &       -       \\
18      &       108.34529       $\pm$   0.00094 &       1.54    $\pm$   0.47    &       2.51    $\pm$   1.76    &       2.70    $\pm$   1.37    & - &     1.63    $\pm$   1.24    &       1.75    $\pm$   1.04    &       -       \\
19      &       84.53237        $\pm$   0.00091 &       1.61    $\pm$   0.43    &       0.29    $\pm$   1.42    &       0.72    $\pm$   1.29    & - &     0.18    $\pm$   0.89    &       0.45    $\pm$   0.81    &       -       \\
20      &       90.30412        $\pm$   0.00087 &       1.46    $\pm$   0.45    &       3.90    $\pm$   1.68    &       0.80    $\pm$   1.26    & - &     2.66    $\pm$   1.41    &       0.54    $\pm$   0.88    &       -       \\ \bottomrule
\end{tabular}
}
\tablefoot{The estimated amplitudes and their ratios were obtained by imposing the frequencies extracted from the $q$-band light curve of PG\,1159--035. Reported amplitudes from the K2 observations ($A_\mathrm{K2}$) are also shown.}
\end{table*}

\end{appendix}
\end{document}